\documentclass{emulateapj}
\usepackage{apjfonts}
\usepackage{epsfig}

\shortauthors{Frinchaboy \& Majewski}

\shorttitle{Open Clusters as Galactic Disk Tracers I}

\begin{document}

\title{Open Clusters as Galactic Disk Tracers: I.  Project Motivation, \\ Cluster Membership and Bulk Three-Dimensional Kinematics}   

\author{Peter M. Frinchaboy\altaffilmark{1,2,3} and Steven R. Majewski\altaffilmark{2}}
\affil{Department of Astronomy, University of Virginia, \\
P.O. Box 400325, Charlottesville, VA 22904-4325, USA}
\email{pmf@astro.wisc.edu, srm4n@virginia.edu}

\altaffiltext{1}{Current Position: National Science Foundation Astronomy and Astrophysics Postdoctoral Fellow, 
University of Wisconsin--Madison, Department of Astronomy, 5534 Sterling Hall, 475 N.\ Charter Street, Madison, WI 53706. }

\altaffiltext{2}{Visiting Astronomer, Kitt Peak National Observatory and 
Cerro Tololo Inter-American Observatory, National 
Optical Astronomy Observatory, which is operated by the Association of Universities for 
Research in Astronomy, Inc. (AURA) under cooperative agreement with the National Science Foundation. }

\altaffiltext{3}{Any opinions, findings, and conclusions or recommendations 
expressed in this material are those of the author and 
do not necessarily reflect the views of the National Science 
Foundation (NSF).}

\begin{abstract}

We have begun a survey of the chemical and dynamical 
properties of the Milky Way disk as traced by open star clusters.
In this first contribution, the general goals of our survey are outlined and the strengths 
and limitations of using star clusters as a Galactic disk tracer sample are discussed.  
We also present medium resolution ($R \sim 15,0000$) spectroscopy of open 
cluster stars obtained with the Hydra multi-object spectrographs on the Cerro Tololo 
Inter-American Observatory 4-m and WIYN 3.5-m telescopes.  Here we use these data 
to determine the radial velocities of 
3436 stars 
in the fields of open clusters within about 3 kpc, 
with specific attention to stars having proper motions in the Tycho-2 catalog.  
Additional radial velocity members (without Tycho-2 proper motions) 
that can be used for future studies of these clusters were also
identified.
The radial velocities, proper motions, and the angular distance of the stars from cluster center are 
used to derive cluster membership probabilities for stars in each cluster field using a non-parametric approach, 
and the cluster 
members so-identified are used, in turn, to derive the reliable bulk three-dimensional motion for 
66 of 71 targeted open clusters.   
The high probability cluster members that we identify help to clarify the color-magnitude 
sequences for many of the clusters, and are prime targets for future echelle resolution spectroscopy 
as well as astrometric study with the Space Interferometry Mission (SIM Planetquest).
\end{abstract}

\section{ INTRODUCTION }


\subsection{Galactic Kinematics Using Open Clusters}

Open star clusters have long been exploited as tools for understanding 
Galactic interstellar dust \citep[e.g.,][]{trump30a,trump30b,cf87,dutra00},
the age of the Galactic disk \citep*[e.g.,][]{ja82,twarog89,pjm94,be17,chaboyer99,carraro},
the Galactic disk metallicity distribution and age-metallicity relation \citep*[e.g.,][]{twarog80,friel93,friel95,twarog97},
and of course stellar evolution \citep*[e.g.,][]{sandage57,cannon70,mm81,mmm93,kr96,pms02}. 
The value of open clusters as tracers of the local Galactic rotation curve has also long been recognized 
\citep*[e.g.,][]{hron87,sfj95,glush98,lb03,pmfthesis}.  It is in this role as a {\it dynamical tracer}
of the Galactic disk that the present study of open clusters is especially focused.

Star clusters can be effective tracers of the Galactic disk because they offer many advantages
over other tracer candidates.  First, relative to other tracers, star clusters lend themselves more 
amenably to age, metallicity, distance, and velocity evaluation.  Compared to an isolated field star at
the same location in the Galaxy, these quantities are much easier to establish in a star cluster, 
needing only a properly interpreted color-magnitude diagram to establish the first three, while
velocities can also be better determined for a star cluster because: (1)
averaging radial velocity and proper motion data over an ensemble of co-moving stars confers potentially
as much as a $\sqrt{N}$ increase in precision for the bulk motion of the ensemble, and (2)
better distances allow one to translate proper motions 
into transverse velocities more accurately.  
 The supplemental knowledge of age and metallicity of a source confers additional beneficial insights 
 into its proper use 
as a dynamical tracer with respect to, for example, assumptions about orbit shape and asymmetric drift.
Alternatively, one can explore Galactic dynamics
as a function of population age and metallicity if all relevant data are available.

On the other hand, 
there are some complications in the use of open clusters as dynamical tracers.  
The challenge of proper identification of cluster members can present particular hazards.
For example, \citet{frin06} showed that the UCAC stars used by \citet{dias06} to establish the
proper motions of at least two particular clusters ---
Be29 and BH176 --- 
in their exhaustive survey of over 400 systems are too bright and cannot be part of these distant systems.
Difficulties caused by 
inaccurate membership censuses are why continued large-scale observational
efforts are needed before we can be confident in the use of open clusters as Galactic disk tracers, 
particularly for more sparse and more distant systems.

To overcome these types of problems, which are typically associated with small number statistics,
it is desirable to survey large numbers of potential open cluster members.  However, this desire
to achieve the largest possible statistical samples often encourages a dangerous
reliance on compilations of disparate data.  
For example the dynamical study of the Galactic
disk using open clusters by \citep{hron87}
found that in their compilation of data from the literature $\sim50$\% of the clusters 
with multiple distance measurements had differences greater than one magnitude in 
determined distance modulus and $\sim50$\% of the clusters also had poor RV qualities.\footnote{It 
is worth pointing out that
this is a common problem with other types of Galactic rotation tracers that have been adopted in 
the past, not just open clusters.}

To help overcome both the membership and homogeneity problems that are often a hindrance to
the use of open clusters as tracers of the Galactic disk, 
we present a new survey of open clusters that will not only take advantage of quality radial velocities 
to help discriminate cluster members, but also rely 
on one source of data as much as possible
for each independent cluster parameter (e.g., all photometry from one source, 
all RVs and derived in a uniform manner,
all proper motions coming from one catalog,  etc.). 
A central objective of this study is the use of these clusters to derive global dynamical properties
of the Milky Way disk, with a particular emphasis on the derivation of the full space velocities for 
the target clusters.  In keeping with our philosophy of uniformity of data, and
because large numbers of proper motions presently must come from all-sky astrometric surveys, 
we first investigate the main proper motion catalogs available for investigations of open cluster 
kinematics and from which we might draw an initial target sample.

\subsection{All Sky Proper Motion Surveys: Hipparcos, Tycho-2, UCAC-2 and the 4M}\label{pmproblem}

A key advance 
that has propelled a resurgence in the use of open clusters as disk dynamical tracers 
is the compilation of all-sky proper motion surveys.
There have been four surveys to determine bulk Galactic cluster kinematics by averaging the proper motions 
of presumed cluster stars based on data from the Hipparcos \citep*{baum00}, Tycho-2 \citep*{dias01,dias02a}, 
Four Million Star (4M) \citep{glush96}, and recently the UCAC-2 \citep{dias06} catalogs.  
The ESA Hipparcos mission has provided critical astrometry for two of these proper motion databases via each of Hipparcos'
primary data products:
(1) the Hipparcos catalog of $\sim118,000$ stars ($V \le 11$) with proper motion uncertainties of 1--2 mas yr$^{-1}$ and 
(2) the Tycho-2 catalog of 2.5 million stars ($V \le 13.5$)  with proper motion uncertainties of 1--3 mas yr$^{-1}$.
The UCAC-2 catalog of 48 million stars is based on Tycho-2 and fainter, ground based observations (to $R = 16$)
and has proper motion uncertainties of 1--7 mas yr$^{-1}$.  The 4M catalog \citep*{4m} was compiled from the Astrographic Catalog and the 
{\it Hubble Space Telescope} guide star catalog (GSC) reduced to the older system of 
the PPM \citep{PPM1} survey, with proper motion uncertainties of $\sim 10$ mas yr$^{-1}$.

\begin{figure}\epsscale{1.0}
\plotone{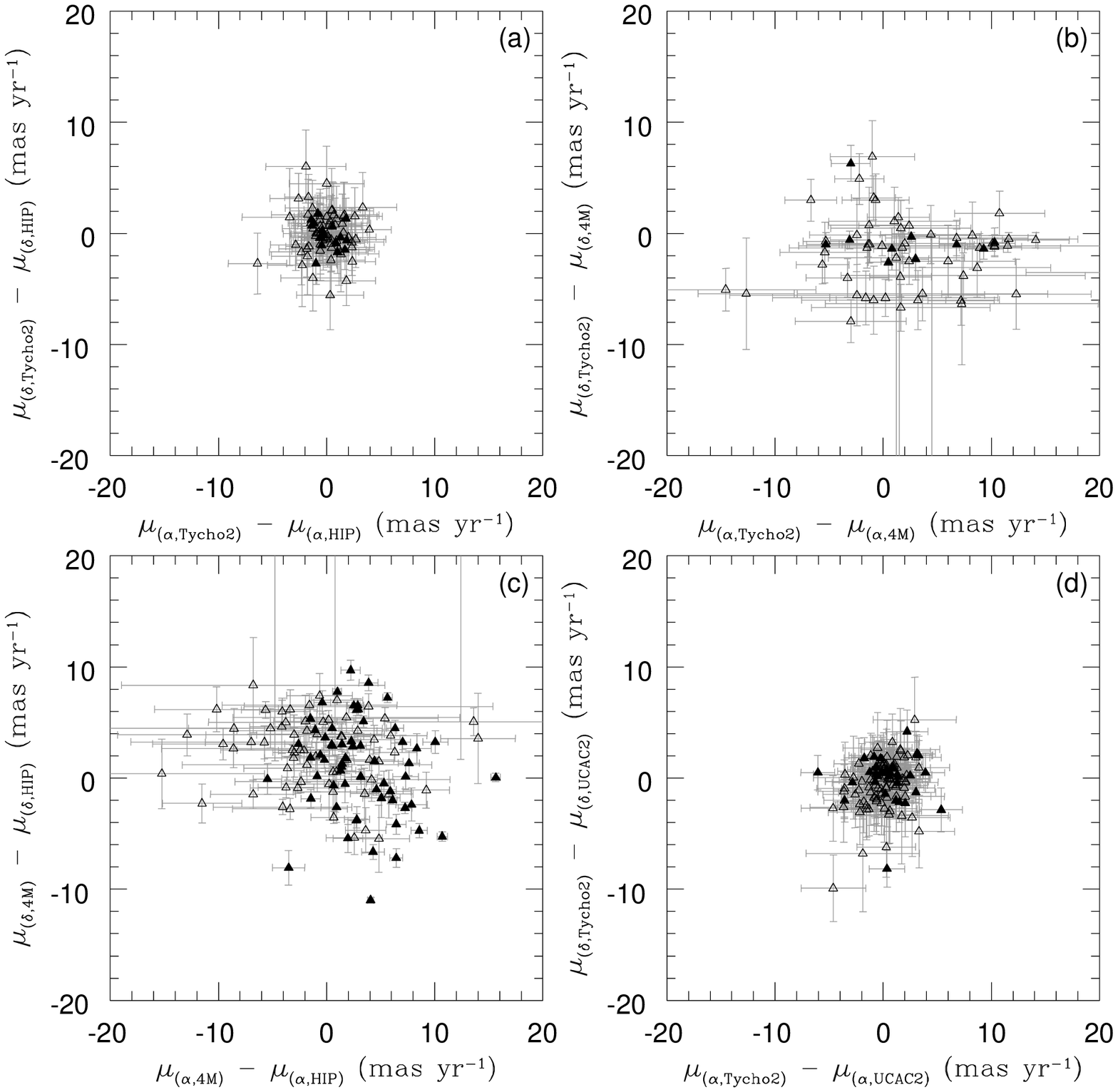}
\caption[Comparison of earlier open cluster proper motion surveys.]{\label{pmvpm}
Comparison of proper motions $\Delta \mu_{\alpha \cos \delta}$ and $\Delta \mu_{\delta}$ derived from the Hipparcos \citep{baum00}, 
Tycho-2 \citep{dias01,dias02a}, 4M \citep{glush96} and the new UCAC-2 \citep{dias06} surveys.  
(a) Hipparcos vs.~Tycho-2. 
(b) 4M vs.~Tycho-2. 
(c) Hipparcos vs.~4M. 
(d) UCAC-2 vs.~Tycho-2. 
Error bars are the quadrature combination of the uncertainties in the two surveys. 
Filled triangles denote clusters with best erors ($\Delta \epsilon_{\mu_{\alpha}}$ 
and $\Delta \epsilon_{\mu_{\delta}}  < 2.0$ mas yr$^{-1}$).}
\end{figure}

Proper motions for hundreds of open clusters have been published using these catalogs.  
However, a comparison of the 
derived motions for the clusters in common between surveys
reveals substantial discrepancies, as shown in Figure \ref{pmvpm}, where 
a cluster by cluster comparison of the proper motion differences is illustrated, namely for
(a) Tycho-2 versus Hipparcos, 
(b) Tycho-2 versus 4M, 
(c) 4M versus Hipparcos, 
and (d) Tycho-2 versus UCAC-2. 
As may be seen, differences in derived proper motions typically 
exceed the quoted uncertainties claimed by each survey.
The best correlation of derived proper motions is between surveys using the Tycho-2 
and Hipparcos astrometry; remaining differences in derived mean cluster proper motions 
between these surveys must therefore be due to differences in the adopted samples of 
presumed cluster members because the actual proper motions, 
at least for $V \lesssim 11$, are the same (i.e. HIPPARCOS-based), 
while the Tycho-2 astrometry used at fainter 
magnitudes is on the Hipparcos reference system (which means that the system is 
referenced to background, extra-galactic sources of the International Celestial Reference System).
Of course, with its bright magnitude limit, Hipparcos
can usually provide useful astrometry for only a small number of stars per cluster 
(typically less than four).  With only a few stars per cluster, a Hipparcos-based survey is far 
more susceptible to small number statistics as well as the misidentification of true
cluster members against the large number of fore/background stars of the Galactic disk.

\subsection{A Closer Look at the UCAC-2 and 4M Catalogs}\label{ucacdescrip}
 
Clearly Tycho-2 and Hipparcos, which are currently the most accurate all-sky astrometric surveys, 
must be considered primary and important sources of proper motion data for our survey.
On the other hand, deeper catalogs can provide more cluster members, but typically with 
worse precision.  Thus, it is not immediately obvious that adding additional data from the deeper proper 
motion catalogs improves or degrades those from Tycho-2 and Hipparcos alone.
A reasonable correlation of derived cluster motions is found when either
the Tycho-2 and UCAC-2 catalog data are used with a $V \le 13$ limit, 
but this is because UCAC-2 adopts Tycho-2 proper motions for stars brighter than about
$V = 13$ \citep{ucac2}.
On the other hand, it is clear that there are greater deviations in derived proper motions when 
we incorporate the fainter stars from UCAC-2. 
Apart from not knowing whether these differences reflect systematic problems in the fainter 
UCAC catalog (which allows probing of proper motions with stars to $V=16$)
or small number statistics in the brighter surveys, some additional concerns about 
UCAC beyond those suggested by Figure \ref{pmvpm}  have led us to not adopt this dataset for our own work.

\begin{figure}\epsscale{1.0}
\plotone{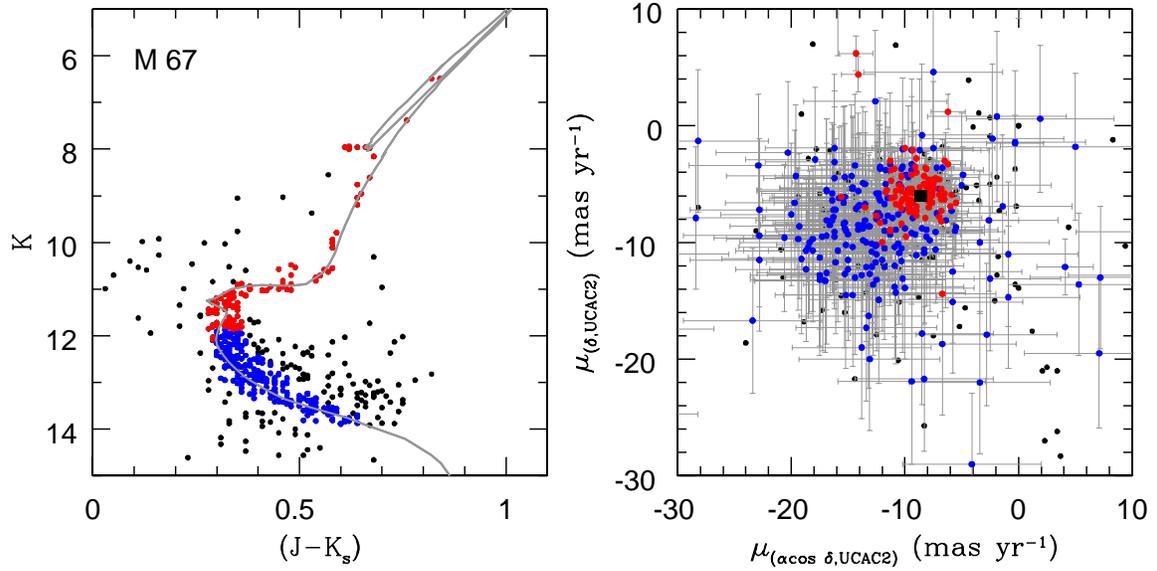}
\caption[Comparison of M67 UCAC-2 data.]{\label{ucacM67}
(a) 2MASS color-magnitude diagram (CMD) of all UCAC-2 stars within the 25' radius of M67 with \citet{sal00} 2MASS isochrone overplotted.
Red points denote stars brighter than magnitude = 13.0 in the UCAC system
(approximately equal to the Cousins $R$ band),
 that are selected to be 
along the cluster's stellar sequence in the CMD
while blue points denote stars fainter than 13.0 
that lie the cluster's main sequence.   
(b) Comparison of proper motions $\mu_{\alpha}$ and $\mu_{\delta}$ derived for M67 stars from the UCAC-2 survey.  
The red and blue points denote the same stars as in (a). One can see that by adding the fainter UCAC-2 data, 
and thereby changing which survey the proper motion data are primarily derived from, 
one can actually change the derived bulk proper motion by almost 2 mas yr$^{-1}$ in each direction.  
The black square denotes the measured bulk proper motion from the \citet{dias01} survey.
}
\end{figure}

For example, as recently pointed out in \citet*{baln}, the UCAC-2 proper motions may have
systematic trends with magnitude due to the compiled nature of the UCAC-2 survey 
(i.e., ground based proper motions are added to Tycho-2 data).
This concern is 
usefully illustrated by looking at the cluster M67 (NGC 2682).  In Figure \ref{ucacM67}a, we show the 
2MASS color-magnitude diagram (CMD) for the M67 field plotting only the most probable members based on CMD location.  
We split the CMD (red: UCAC mag $< 13.0$, blue:  UCAC mag $\ge 13.0$) at a magnitude
that represents approximately the transition within the catalog from Tycho-2 to 
ground-based observations.
Substantial proper motion shifts are apparent between the bright and faint samples 
(Figure \ref{ucacM67}b), and this suggests significant systematic zero-point offsets within 
the UCAC-2 database.  

The 4M catalog, as well, appears to have systematic proper motion errors.  The 4M is
not tied to the Hipparcos system, but rather to the PPM \citep{gn92}. 
\citet{glush96,glush98} have used the 4M catalog to determine the proper motions
of about 200 open clusters.  \citet{dias01} compared their Tycho-2 open cluster 
proper motions to both those based on Hipparcos \citep{baum00} and the Glushkova et al. 4M work
and found that the 4M motions were systematically
offset from those in the Hipparcos system by
$\sim 5$ mas yr$^{-1}$ in both 
$\mu_{\alpha} \cos \delta$ and $\mu_{\delta}$, an amount that was larger than 
expected given the quoted errors of both the surveys.
While these differences likely reflect both differences in membership as well as astrometric
accuracy, this comparison suggests that
the deeper proper motions are not necessarily providing 
better overall accuracy in the open cluster bulk motions, and recommends a strategy
based on quality over quantity of cluster star motions.

Therefore, because of uncertainty over the reliability of the UCAC-2 and 4M surveys
and our desire to adhere to a ``quality over quantity'' policy, we have elected
to focus on deriving bulk motions using astrometry from  
the Tycho-2 catalog, but with dedication to ensuring
that we derive a trustworthy membership of the smaller number of available cluster stars 
available in this shallower database.


\subsection{A New Galactic Tracer Survey}

The mass and mass distribution of the Galactic disk 
has been a matter of debate for over a century, and will likely remain so until 
extremely precise proper motions and trigonometric parallaxes can be obtained for 
numerous disk tracers, most likely through 
future space-based studies like the National Aeronautics and Space Administration's (NASA) 
{\it Space Interferometry Mission} (SIM PlanetQuest) and the European Space Agency's (ESA) {\it Gaia} satellites.  
The new project presented here represents both a preparatory 
effort in this space-based direction as well as a standalone dynamical study in its own rite.  Our goal is to
establish a well-constructed, well-studied, baseline tracer population --- open clusters --- that can not
only (1) serve as input targets for Galactic dynamics studies with 
SIM PlanetQuest (specifically, for the SIM Key Project
{\it Taking Measure of the Milky Way}, for which SRM is the Principal Investigator and which
has provided support for this project), but 
which (2) can also be immediately exploited for understanding Galactic dynamics with existing
astrometric data.

As mentioned above, 
the inability to establish a uniform, unbiased tracer sample has been one of the key 
weaknesses of previous Galactic dynamical surveys \citep*[e.g.,][]{fbs89}.
To provide a homogeneous set of tracers, 
we have undertaken a spectroscopic survey to obtain
precision RVs of open cluster fields.  These RVs will establish cluster membership for individual stars
that not only provides a very precise mean RV of each cluster, but, in identifying cluster members 
having accurate astrometry, can be used to define the bulk cluster proper motion.  
The combination of the newly found, very precise mean RV of each cluster with its
derived bulk proper motion and distance will
allow us to determine the space velocities of these clusters.
With a large number of cluster space velocities, the rotation curve of the Galactic 
disk can be constrained over the $R_{gc}$ 
range of the sample.  Alternatively, through the adoption of an assumed rotation curve 
(i.e., Galactic potential), the orbital properties of individual clusters can be determined.

Because we are interested in obtaining results before SIM PlanetQuest and Gaia
are in service, our RV study will focus on clusters already having available, uniform
and reliable proper motions.  
As described in \S1.4, 
we have elected to focus on the all-sky Tycho-2 proper motion catalog, 
which provides useful astrometry for
typically 50--200 stars per 
cluster field ($\le 0.75$ deg$^2$).  While selection of Tycho-2 as our source of proper motions
limits the depth and thereby the cluster distance that can be 
explored, it is in keeping with our philosophy of quality over quantity for the astrometric data.
The selected proper motion 
stars for a given cluster field can usually be investigated spectroscopically with a single pointing of the 
NOAO Hydra multi-fiber spectrographs on the CTIO 4-meter and 
WIYN\footnote{The WIYN Observatory is a joint facility of the University of
    Wisconsin-Madison, Indiana University, Yale University, and the
    National Optical Astronomy Observatories.} 3.5-meter telescopes, and the radial velocities 
derived from these spectroscopic data are the primary results presented here.
Our campaign of multi-fiber spectroscopy allows us to check virtually every star in a cluster field 
having a Tycho-2 proper motion, and leaves additional fibers
to (1) expand the RV membership census to fainter stars in anticipation of the future astrometric
surveys (e.g., SIM and GAIA), and (2) improve age-dating of the clusters through CMD-isochrone 
fitting to established member stars. 
The current study of clusters provides a large uniform database for further open cluster research, 
as a supplement to the \citet{dias02b} and WEBDA \citep{webda} databases.

The new RVs immediately improve all previous proper 
motion work on our targeted clusters because of the clarity they bring regarding cluster membership.  
The improved RVs and proper motions, when combined with new distances we shall 
derive elsewhere (Paper II), will provide
much more reliable space motions of numerous open clusters over a large $R_{gc}$ range; these
space velocities will be
at a precision sufficient to make tangible improvements in the determination of the nearby 
Galactic rotation curve and, in turn, the mass 
distribution of the Galactic disk.  With uncertainties of order $\sim1.2$ km s$^{-1}$,
the data here yield the best derived bulk RVs thus far for most of the chosen clusters, 
This precision is comparable to the uncertainties in transverse velocity that SIM and Gaia
will measure for these clusters, and represent a significant 
improvement over many  previous RV surveys of open clusters, which have typical uncertainties
of order $\sim15$  km s$^{-1}$ \citep{sfj95}.  Our results are more comparable to the RV precisions
being obtained for open cluster stars in studies using CORAVEL \citep[e.g.,][]{mm89,mm90}, for example.

Following the work in this contribution (Paper I), 
we will provide uniformly-determined distances and ages derived from isochrone-fitting to
2MASS photometry of these clusters,
aided by the cluster membership data derived here (Paper II).  
With newly-derived kinematics and distances in hand from Papers I \& II, 
we will then use the cluster sample to explore not only the orbital characteristics of the
individual clusters (Paper III), but global properties of the Galactic disk (Paper IV), including:
(1) the local Galactic rotation curve and velocity field near the Sun,
(2) the kinematics of the disk across the frontier separating $R<R_0$ and $R>R_0$, and 
(3) the validity of the assumption of Galactic dynamical symmetry 
(e.g., north vs.\ south, Galactic quadrants I/II vs. IV/III).

In \S6 and Table 12 of this paper we present the 
derived 3D space motions of the clusters that enable these future contributions.
In the preceding sections of this paper we explain how we selected our target clusters (\S2.1 and \S4) 
and which stars within each cluster field to probe (\S2.2), the spectroscopic observations and the derivation 
of radial velocities (\S3), and the means by which membership within each cluster is established
(\S5).


\section{Source Selection}

\subsection{Cluster Sample Selection}
    
Our selection of specific open clusters starts with the 205 clusters explored in the \citet{dias01,dias02a} catalogs, 
which derive cluster membership using 
the statistical method of \citet{sand77}.  
We also adopt the following criteria: (1) the clusters must have 
at least ten stars with Tycho-2 proper motions in the fields selected by \citet{dias01,dias02a}, and
(2) the cluster diameters cannot be much larger than the Hydra field of view 
(40\arcmin -- CTIO, 60\arcmin -- WIYN) 
so that the cluster can be sampled with a significant number of fibers.
In addition, to obtain the greatest leverage on the local Galactic rotation curve
the selected clusters span a wide area over the 
Galactic $X_{gc}$-$Y_{gc}$ plane and reach
to a heliocentric distance of $\ge$2.5 kpc.
Neither age, distance from the Galactic plane, nor metallicity was considered as a selection criterion.

\begin{figure} \epsscale{1.01}
\begin{center}
\plotone{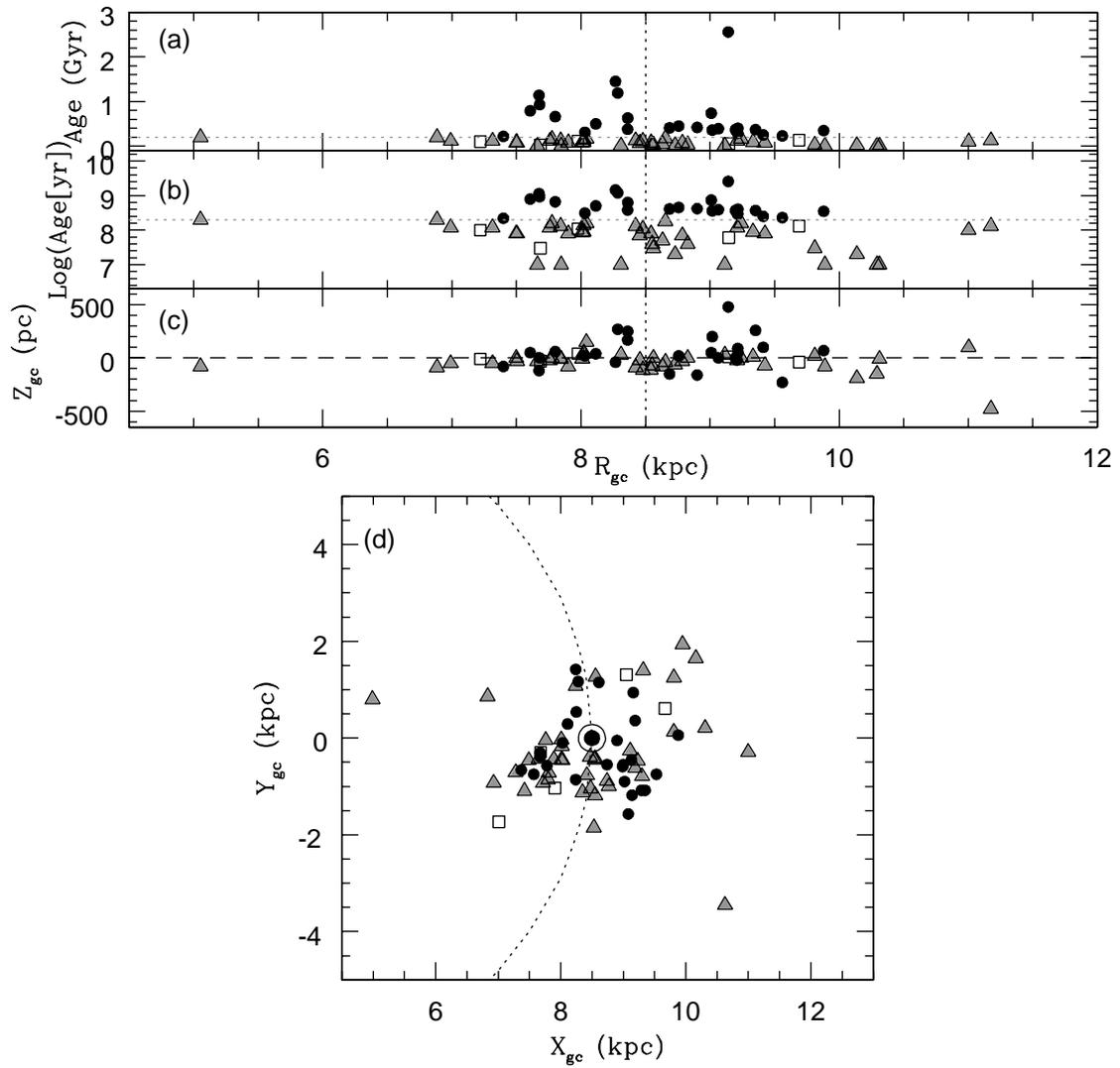}
\end{center}
\caption[Parameters of the Open Cluster Sample]{\label{CHARplot} Properties of the open cluster sample.
(a) Plot of 
cluster age vs. $R_{gc}$.  One can see that most clusters are less than 200 Myr old (grey triangles) 
though a number of old clusters are present in the sample (black circles).  Open Squares denote the clusters Collinder 258,
Lynga 1, NGC 1513, NGC 6250, and NGC 7654 (see \S \ref{spacemotions} \& \ref{comp2dias}), 
which we excluded from our sample
because our results for these clusters are uncertain.
(b) Same as (a), but showing a smaller age range.
(c) Distribution of $Z_{gc}$ (height above/below the Galactic plane) versus $R_{gc}$.
(d) $X_{gc}$, $Y_{gc}$ distribution of clusters in this study. } 
\end{figure}

Table \ref{cltargets} shows the basic cluster parameters of our sample with data taken
from the \citet{dias02b} catalog, including coordinates of 
right ascension and declination (cols. 2 and  3) and Galactic longitude and latitude (cols. 4 and 5), 
heliocentric distance (col. 6),
$\log({\rm age/years})$ and visual diameter of the cluster in arcminutes (cols. 7 and 8), 
and the observing run on which the cluster was observed (see below and Table \ref{crvsv} 
for definitions).
The Galactic distribution of our final cluster sample of 71 clusters
 is shown in Figure \ref{CHARplot}.  
The smaller number of clusters we have sampled in the $l = 0$--180\degr~half of the Galaxy is result of a smaller amount of 
observing time obtained for the WIYN observations; however future work in the research program will aim to 
remedy this deficiency.
Figure \ref{CHARplot} also shows the distribution of the selected cluster ages and distances from the
Galactic midplane as 
a function of their Galactic radius (assuming the Sun is at 8.5 kpc).
More than half of our final sample have ages less than 200 Myr but older than 10 Myr
(Table 1).
The large number of relatively young clusters is 
important for kinematical studies of the Galactic disk because, in general, open clusters should 
develop increasing deviations from ``normal'' 
disk rotation due to the scattering by molecular clouds over time \citep{ss51,ss53}; 
however, 
clusters that are {\it too} young may still reflect
the specific dynamical environment of their birth and may not yet
have circular orbits \citep{lyngaP87}. 
Figure \ref{CHARplot}c shows that all clusters in our sample are within 500 pc 
of the Galactic plane, and most are within 200 pc.
This supports the notion that most clusters in our sample are likely to be
 ``well behaved'' in the sense that they have not been scattered far from the 
 Galactic midplane and therefore are likely to still be on near-circular orbits (of course, we 
 will revisit this question when we examine cluster orbits in detail, in a future contribution).

Less than 25\% of our clusters have estimated metallicities ([Fe/H]), 
so we have little leverage on this aspect of our sample; 
however, we hope to derive metallicity estimates for some of our clusters in the
future, using not only improved isochrone fits to CMDs aided with our membership data, 
but the spectra themselves.


\subsection{Stellar Selection Within Each Cluster}\label{targselect}

Given that bulk 3D motions are our primary goal, 
the first stellar targets within each cluster selected for observation 
were those Hipparcos and Tycho-2 stars used in the \citet{dias01,dias02a} survey for the clusters.  
Because constraints on fiber optic placement with the Hydra instrument 
(i.e., two fibers cannot be closer than 25\arcsec~in the Hydra setup) 
mean that in some cases not all desired stars can be 
observed in a cluster field, we must prioritize stars within a fiber setup.  
For this reason, stars were ranked in priority order based on the \citet{dias01,dias02a} derived membership probabilities, 
from highest to lowest probability.  
\citet{dias01,dias02a} derived these probabilities based on the proper motions using the method 
of \citet{sand77}.

Next, additional Tycho-2 stars available in the Hydra field of view, but not used in 
the \citet{dias01,dias02a} study (because they lie beyond the cluster radius studied by these authors)
were added as the next priority to the target list.
For the WIYN/Hydra runs, no targets beyond the Tycho-2 stars needed to be selected because 
the combination of a smaller number of available Hydra fibers (90 vs. 132 for CTIO/Hydra) 
and larger field of view (60\arcmin~vs. 40\arcmin~for CTIO/Hydra) 
typically meant that nearly all target fibers were filled with Tycho-2 stars.

For the CTIO runs and for fields having less than 50 stars with available Tycho-2 proper motion data, 
we selected at lowest priority two additional sets of stars; first,
stars between $V=13$--15 magnitude from the USNO-B1.0 catalog from within 
the cluster radius 
(with that value taken from the Dias et al.\ 2002b catalog), 
 with the goal of searching 
 for additional cluster members fainter than the $V \sim 13.5$ magnitude limit of the Tycho-2 survey, and
second,  we allowed unused ``field orientation probe stars'' (FOPS; USNO B1.0 stars with $12 < R_2 < 13$) 
to be added to the bottom of the target priority list. 
At either WIYN or CTIO, fibers that were not assigned to targets were used for sky observations, 
with at least six (WIYN) or ten (CTIO) fibers positioned on random sky
for sky subtraction of the stellar spectra.


\section{Spectroscopic Survey Data}
\subsection{Spectroscopic Observations}

We have collected homogeneous spectroscopic observations for 71 open clusters 
using the HYDRA multi-fiber spectrographs on the Blanco 4-m telescope 
at Cerro Tololo Inter-American Observatory (CTIO) and the 3.5-m WIYN\footnote{The WIYN Observatory is a joint 
facility of the University of Wisconsin-Madison, Indiana University, Yale University, and the
National Optical Astronomy Observatories.} telescope 
at Kitt Peak National Observatory (KPNO). 
This project was conducted using publicly competed NOAO time and was granted long-term 
status\footnote{This project was selected as an NOAO Ph.D. thesis project for PMF.}, which permitted 
observations in the semesters 2002A--2004A over a total of fourteen awarded CTIO 4-m and 
six WIYN 3.5-m nights.  The data for the clusters listed in Table \ref{cltargets} were obtained the 
nights of UT 2002 March 8--12 (``Run 1''),  
2003 March 16--21 (``Run 2''), 
2003 July 20--23 (``Run 3''), and 
2003 August 2--8 (``Run 4'') from CTIO.  
For more efficient observing, some cluster observations scheduled for two August 2003 CTIO 
nights were interspersed with other targets for two other observing projects awarded telescope
time 
over the course of eleven CTIO/Hydra nights in July and August 2003 (Runs 3 and 4). 
The WIYN data were observed on the nights of 2003 September 14--18 (``Run 5'').

The CTIO observations made use of 132 Hydra fibers that are simultaneously 
dispersed onto a $2048\times4096$ pixel, SITe400mm
CCD using the 380 grating with 1200 lines mm$^{-1}$ and with the fiber ends viewed 
by the spectrograph through the 100$\mu$m slit plate to improve the resolution to
a dispersion of 0.68 \AA ~per resolution element ($R \sim 15,000$).  The spectral range covered 
was 7740--8740 \AA.
Data obtained at WIYN dispersed the 90 Hydra fibers dispersed onto a $2048\times2048$ pixel CCD 
in the Red Bench Camera 
using the echelle (316@63.4) grating in 6th order; this yielded 
a dispersion of 0.82 \AA ~per resolution element ($R \sim 13,000$) which was centered on the 
8220--8800 \AA\  spectral range.
Typical signal-to-noise ratios ($S/N$) of 10 or better were obtained;
all cluster stars presented have at least $S/N \ge 5$.  
To aid the RV calibration, multiple RV standards were observed on each
run, where each ``observation'' of an RV standard entails sending the light of the calibrator
down 2-12 different fibers, yielding many dozen individual spectra of each RV standard.  
We present here the results from analysis of spectra for 3436 individual stars out of
3537 with sufficient $S/N$ observed in the fields of 71 open clusters.
\footnote{Of the 3537 stars
observed with $S/N > 5$, 101 were peculiar stars (e.g., carbon stars, Be stars, young emission-line stars) 
that are excluded from the RV analysis using ``normal star'' cross-correlation 
templates discussed here.}

\subsection{Data Reduction}\label{data_redc}
Preliminary processing of the two-dimensional data was undertaken using standard $IRAF$\footnote{IRAF 
is distributed by the National Optical Astronomy Observatories, which are
operated by the Association of Universities for Research in Astronomy,
Inc., under cooperative agreement with the National Science Foundation.}  
techniques as described in the $IRAF$ {\ttfamily ccdproc} documentation.
After completing the CCD bias subtraction, overscan
correction and trimming, the two-dimensional images were corrected 
for pixel-to-pixel sensitivity variations and chip cosmetics
by applying ``milky flats'' 
according to the prescription outlined in the CTIO Hydra manual by
N. Suntzeff.\footnote{http://www.ctio.noao.edu/spectrographs/hydra/hydra-nickmanual.html}

After basic processing the data were run through the $IRAF$ routine {\ttfamily dohydra}.
One dimensional spectra for each star were extracted from the two-dimensional CCD images 
and wavelength calibrated with respect 
to a comparison lamp spectrum.  Exposures of the PENRAY (CTIO; He, Ne, Ar, and Xe) or CuAr (WIYN) lamps were taken 
at each Hydra pointing through all fibers to provide comparison spectra yielding at least ~11 
prominent emission lines roughly evenly distributed over the observed wavelength range. 
These comparison spectra provide a wavelength solution (i.e., pixel to wavelength conversion) for each 
extracted object spectrum. 

\subsection{Stellar Radial Velocities: Standard Stars \label{stddescrip}}

All radial velocities were derived using IRAF's {\ttfamily fxcor} package, which 
we used first to determine RVs for the standard stars.
Radial velocity standard stars are used to check for systematics in the data, 
to determine the measured RV precision, and to calibrate the zero-point of the velocity scale.
The reduction to radial velocities employed essentially the classical
cross-correlation methodology of \citet{td79}.  
The template star input to the correlator is prepared from a high signal-to-noise ($S/N$) 
standard star spectrum from which the stellar continuum is fitted and subtracted.   
The resulting spectrum is high and low pass Fourier-filtered to remove both high frequency noise (e.g., cosmic rays) 
and the low frequency variation cause by difference in instrument throughput.

We first measured the RVs of standard stars by cross-correlating each Fourier-filtered standard star
spectrum against every other standard star spectrum from its corresponding observing run 
over the restricted wavelength range of 8220--8680 \AA, which avoided possible contamination from nearby atmospheric lines.
The resulting RVs from different cross-correlations for each individual standard star spectrum were 
averaged and the standard deviations measured; the results are presented in Table \ref{rvstdindiv}. 
The average velocity standard deviation for the individual Hydra standard star spectra is 
$\sigma_v \lesssim 2$ km s$^{-1}$ for spectra with $S/N \ge 20$.

We determined the level of the random and any unknown systematic RV errors from a prescription described in \citet{vogt95}, which 
is based on the analysis of repeatedly observed stars (Table \ref{rvstdindiv}).  In this
case these stars were typically observed through different Hydra fibers.
The Tonry--Davis Ratio \citep[TDR;][]{td79} for each spectrum is measured using {\ttfamily fxcor}.  
Since the TDR scales approximately with $S/N$ we can, following the method described in \citet{vogt95},
determine approximate 1$\sigma$ errors in the RVs corresponding to a given TDR as:
\begin{equation}
\label{erroreq}{\rm error}~V_{r} = \frac{\alpha}{(1+{\rm TDR})}.
\end{equation}
The parameter $\alpha$ is a constant calibrated by the standard star data using the following formula, 
which is predicated on the assumption that the TDR is a good measure of the relative $S/N$, and
where autocorrelations are not included:
\begin{equation}
\label{erroreq2} \alpha^2 = \frac{\sum_i\sum_j(1+{\rm TDR}_{i,j})^2(V_{r,i,j}-\langle V_{r,j}\rangle)^2}
{\chi^2_{50,n}}
\end{equation}
\noindent where $V_{r,i,j}$ is $i$th observation of the $j$th standard star, and  $\langle V_{r,j}\rangle$
is the mean RV of the $j$th standard star.
We obtained the values of $\chi^2_{50}$ for our sample's number of degrees of freedom,
where $\chi^2_{50,n}$ is the critical value of the $\chi^2$ distribution 
at the 50\% confidence level multiplied by $n$ degrees of freedom, as described fully in \citet{vogt95}.

Since our target stars were selected based on proper motion criteria, they span a wide range of spectral types, 
including anything from hot O and Be stars to cool carbon stars. As a result, we observed a range of 
RV standard star templates.
However, due to the lack of International Astronomical Union (IAU) RV standards hotter 
than spectral type A0, we used B and A stars 
from \citet{fekel} to provide RV standards for our hot star spectra.
For a better match in the spectral types between targets and cross-correlation templates,   
we divided the observed standard stars into ``red'' or ``blue'' subsamples for our cross-correlation templates.
Stars were 
considered ``red'' stars if the \ion{Ca}{2} infrared triplet (8498\AA, 8542\AA, and 8662\AA) 
was present in the spectra; this encompasses cool F through early M type stars.   
``Blue'' stars have dominant Paschen series lines and virtually  
no \ion{Ca}{2} triplet (i.e., O through A type stars).

Initially we anticipated using primarily late type stars for our analysis, so that in the earlier runs 
(March 2002, March 2003 and July 2003 
observations; Table \ref{cltargets} -- Runs 1, 2, 3) we did not obtain ``blue'' standards.
Later it became clear that good RVs for hotter stars could be derived and we began to collect blue standards.
To cover the lack of blue standards in the earlier runs, the ``blue'' August 2003 standards were used to reduce 
all CTIO ``blue'' target stars.  This approach was adopted and found to work 
moderately well 
because even across observing runs all spectra were taken with the same instrument setup, 
are dispersion corrected uniformly, and should
experience no flexure problems because Hydra uses a bench-mounted spectrograph.  
In the end we did find some offsets in the RV zero-points for some of the runs --- but, ironically
these were for the runs where we actually did take blue standards (see Sections 3.6 and 3.7).

The standard star spectra (Table \ref{rvstdindiv}) provide a data set for calibration of the RV errors 
for each run according to Equations 1 and 2.
Table \ref{rvstdindiv} lists each observation of a standard star, the UT date of the observation (col. 2), its spectral type (col. 3), 
which observation frame and fiber were used (cols. 4 and 5), the TDR (col. 6), 
the mean derived radial velocity ($V_r$) and the measured standard deviation (cols. 7 and 8), followed by the average $V_r$ and 
standard deviation for the cluster (cols. 9 and 10), also
the IAU or \citet[][]{fekel} RV is shown for comparison  in col. 11.
Using the \citet{vogt95} technique and measurements from Table \ref{rvstdindiv}, 
we find the parameters given in Table \ref{crvsv}, which are used to determine errors 
for each of the runs.  The variation in the $\alpha$ values can be due to the effects of 
focus and small spectrograph setup variations.  


\subsection{Radial Velocity Standard Verification}

\begin{figure}
\begin{center}
\epsscale{0.55}
\plotone{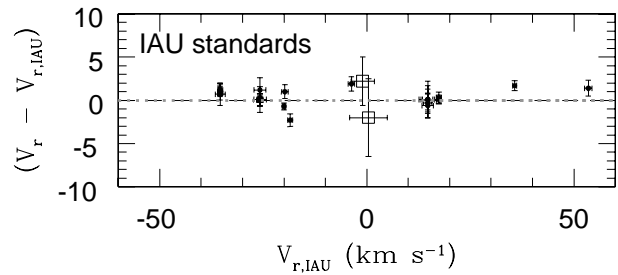}
\end{center}
\caption[RV comparison for RV standard stars]{\label{RVcompIAUplot} Comparison of the
measured RVs for \citet[][ open boxes]{fekel} and IAU radial velocity standard stars.
The dashed line marks an ideal 1-to-1 correlation.  The dotted line is a linear fit to the data.
}
\end{figure}

To check the reliability of our measurements of radial velocity standard stars, 
we compare the difference between our measurements  (Table \ref{rvstdall})
and the IAU or \citet[][]{fekel} values, as shown in Figure \ref{RVcompIAUplot}.
As may be seen, the derived RVs for the standard stars are
all within 2.3 km s$^{-1}$ of the IAU values.  
We find that the difference between our measurement and the IAU values are no more than 2 times larger than the quadrature errors 
(as shown in Table \ref{rvstdall}
); however, we find that the differences are randomly distributed 
and that the mean offset is less than 1 km s$^{-1}$.
Therefore we find there are no 
systematic trends between our measured velocities and the cataloged values for the IAU standards, though we did find an offset for 
the ``blue'' \citet{fekel} stars, a situation that is analyzed in more depth below (\S \ref{RBchk}).

\subsection{Stellar Radial Velocities: Target Stars}
\begin{figure}\epsscale{1.0}
\epsscale{1.0}
\plotone{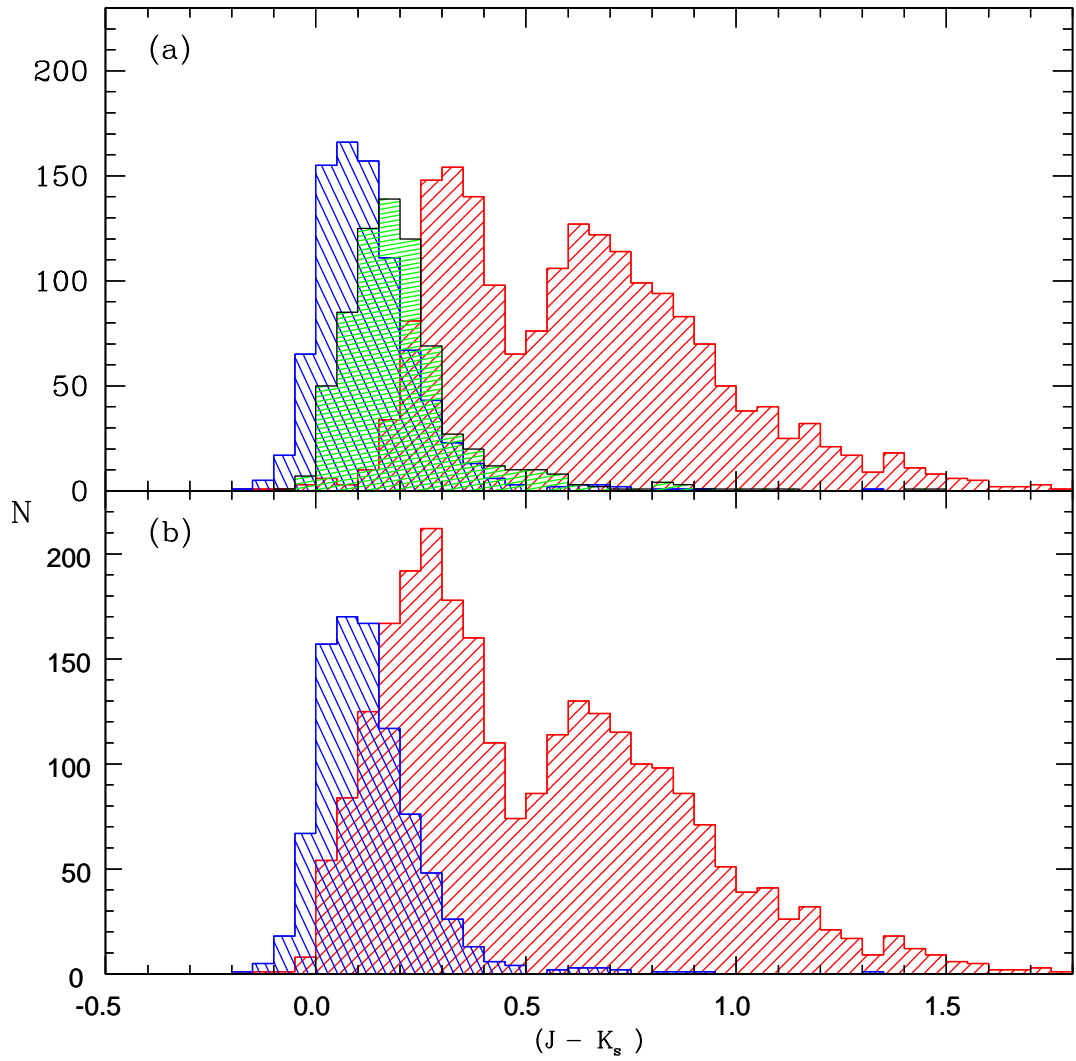}
\caption[Comparison of Template selection]{\label{tempVcolor} 
Comparison of the color/temperature of the cluster field stars to the selected RV template.
(a) Raw selection of stars into ``red'', ``green'', and ``blue'' based on the appearance of the spectra.
(b) The distribution of stars cross-correlated vs. IAU RV standards (red) and \citet{fekel} RV standards (blue).
}
\end{figure}

Target stars were analyzed using the same reduction procedure as the standard stars, with the exception that 
stars that showed both the Paschen series lines and any hint of the \ion{Ca}{2} triplet were considered ``green'' stars.
As with the RV standards, the targets were sorted into ``red,'' ``blue,'' and ``green'' sub-samples based on 
visual inspection of their spectral features in order to match them to the appropriate cross-correlation template.
Figure \ref{tempVcolor}a shows the 2MASS color distribution of the stars selected for each sub-sample.  
The ``green'' stars were tested against both templates to find the best match.  Nearly all ``green'' stars became part of the ``red'' sample.

Each group of target stars was processed through IRAF's {\tt fxcor} package to cross-correlate them against the standard of the
corresponding color class for their respective observing run, as described in \S \ref{stddescrip}.
``Green'' stars were cross-correlated against 
both red and blue templates and the derived RV was taken from the template that provided the better result. 
The final 2MASS color distribution for stars fitted with the ``red'' and ``blue'' templates is shown in  Figure \ref{tempVcolor}b.
Uncertainties for the stars fitted to the ``red'' or ``blue'' templates
were determined using the $\alpha$ values from Table \ref{crvsv} and Equation \ref{erroreq}.


\subsection{Internal Comparison (Red vs. Blue)} \label{RBchk}

\begin{figure} 
\begin{center}
\epsscale{1.0}
\plotone{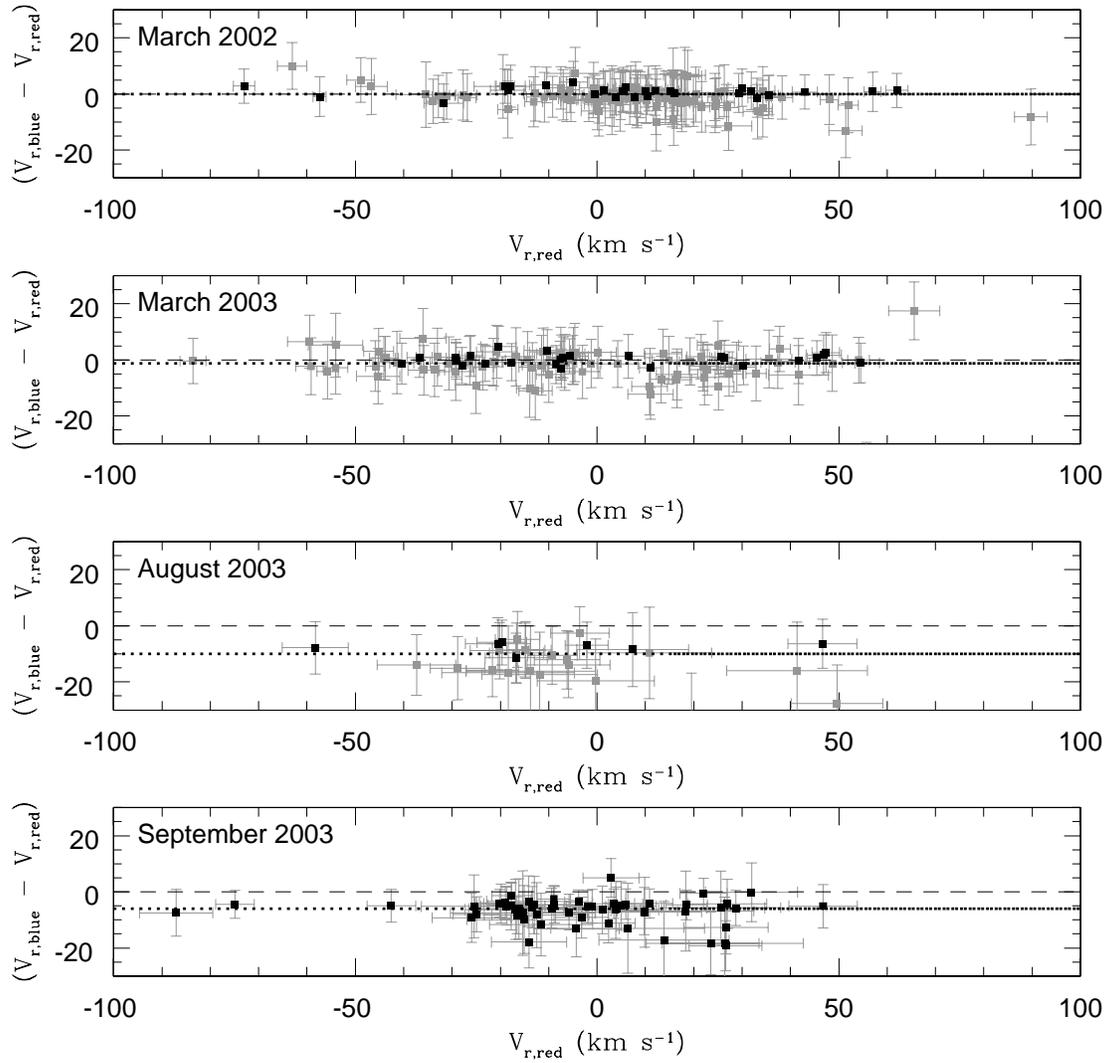}
\end{center}
\caption[Red vs. Blue RV comparison for all runs]{\label{RVcompRBplot} 
Comparison of RVs derived for the ``green'' spectra (which show both 
the \ion{Ca}{2} triplet and the Paschen series features) using red versus blue
cross-correlation templates.  
All stars shown have RV measurement errors in the blue less than 10 km s$^{-1}$, 
the black squares have RV measurement errors in the blue less than 6 km s$^{-1}$. 
The black dashed line shows a perfect 1-to-1 correlation and the dotted line is a linear fit to the trends.
(a) The March 2002 data shows less than a 1 km s$^{-1}$ shift, and therefore no offset was applied between the red and blue data. 
(b) March 2003 shows a $-$1 km s$^{-1}$ shift, which was applied to the ``blue'' RVs. 
(c) August 2003 shows a $-$10 km s$^{-1}$ shift, which was the correction applied to the ``blue'' RVs.
(d) September 2003 shows a $-$6 km s$^{-1}$ shift, which was applied to the ``blue'' RVs.
Note: the July 2003 run had no stars
where the blue 
template error was less than 10 km s$^{-1}$.
}
\end{figure}

As an additional check on the measured RVs, we tested the internal consistency 
delivered by the separate ``red'' and ``blue'' 
reductions for a given run.  To do this, stars in our ``green'' sample  
were correlated with both the ``red'' and ``blue'' standards.
For stars with measured uncertainties in their blue measurement of less than 10 km s$^{-1}$, 
Figure \ref{RVcompRBplot} shows $(V_{r,blue} - V_{r,red})$ vs.~$V_{r,red}$, where stars with the best
blue uncertainties 
($\le 6$ km s$^{-1}$) are denoted with black squares.
The subsample of stars with uncertainties of $\le 6$ km s$^{-1}$ was then fitted with a line 
to determine if any zero point offset was needed between the red and blue samples.  
The fit to the data in the $V_{r,red}$ vs. $V_{r,blue}$ plane is given by:
\begin{equation}
V_{r,adopted} = a_0 + a_1 * V_{r,blue}. 
\end{equation}
\noindent The resulting fits for the March 2002, 2003, August 2003 and September 2003 runs are given in Table \ref{fitRB}. 
We did not find a systematic difference between the $\le$10 km s$^{-1}$  and $\le$6 km s$^{-1}$
samples, just a larger scatter for the $\le$10 km s$^{-1}$ data.

We find that the blue data are offset from the red data by a significant amount for the 
August 2003 and September 2003 runs, and a small offset is found for the March 2003 run.  
To verify that it is the red values that are more 
reliable and to support the rationale that we shift the blue system RVs to the red system, we next compare our 
RV measurements from the red sample with those of previously published, high RV resolution studies for a number of open clusters.

\subsection{Systematic Effects and Comparison to Previous Results}

As an additional test of the reliability of our RVs, we have found previously published 
values for stars in nine clusters that we have observed.
We undertook a comparison of our velocity measures to those in the following studies: 
IC 4561 \citep*{mermio95,meibom02}, IC 4756 \citep{mm90},
NGC 2099 \citep*{mermio96}, NGC 2423 \citep{mm90}, NGC 2447 and NGC 2539 \citep{mm89}, 
NGC 2682 \citep[M67; ][]{mathieu86}, 
NGC 5822 \citep{mm90}, and NGC 6134 \citep*{mermio92}.
In Table \ref{clM67comp}, we present a direct  star-by-star comparison of RV results 
to the seminal work on M67 by \citet{mathieu86}.
Table \ref{clcomp} provides star-by-star comparison for the other clusters listed above, 
which include the Tycho-2 star name, the star name from the corresponding photometry reference used for identification in the previous RV studies, 
the stellar coordinates, our RV ($V_r$) and its uncertainty, the reference RV and its uncertainty, and the per star difference in these measurements.
For stars in common between the surveys, we find overall excellent agreement in the determined per star kinematics.

\begin{figure} 
\begin{center}
\epsscale{1.01}
\plotone{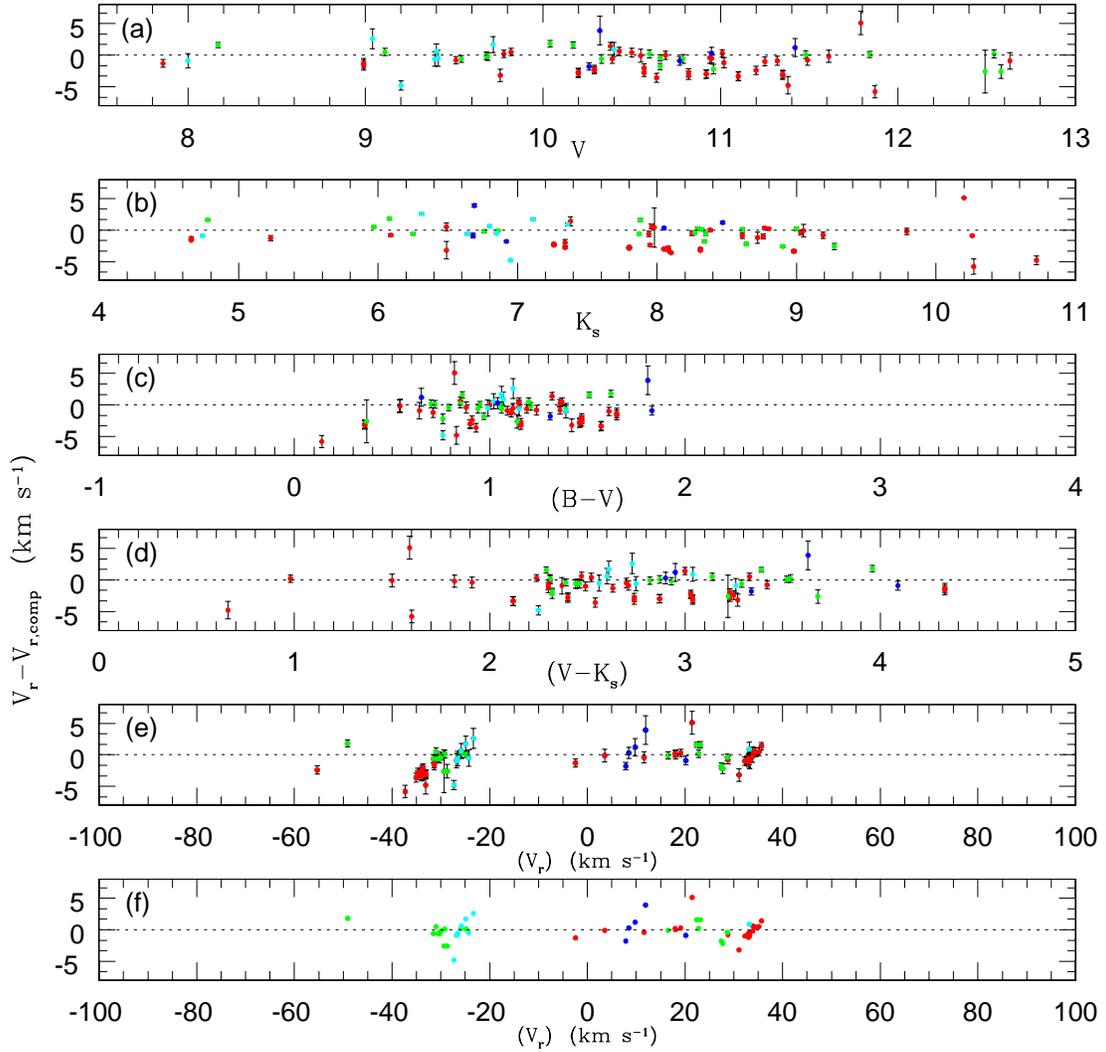}
\end{center}
\caption[Comparison vs. Previously Published Results]{\label{RVcompplot} 
Comparison of our measured RVs versus previously published star-by-star high precision velocity results.
All stars are color-coded by the observing run on which they were observed as follows: red = March 2002; 
green = March 2003; cyan = August 2003; and blue = September 2003. 
(a) $\Delta V_r$ vs.~the $V$ magnitude converted from the Tycho ($V_T$) magnitude.
(b) $\Delta V_r$ vs.~the 2MASS $K_s$ magnitude. 
(c) $\Delta V_r$ vs.~the ($B-V$) color converted from the Tycho ($V_T$,$B_T $) magnitudes.
(d) $\Delta V_r$ vs.~the ($V-K_s$) color, where $V$ is 
converted from the Tycho ($V_T$) and ($K_s$) is the 2MASS magnitude.
(e) $\Delta V_r$ vs.~our measured $V_r$. 
(f) Same as (e) except with cluster IC 4651 removed.
}
\end{figure}

In Figure \ref{RVcompplot}, we compare differences between our own red data
and previously published RVs as a function of 
photometric parameters (e.g., magnitude and color) of the stars, 
where the data are color-coded by observing run 
(red = March 2002, green = March 2003, cyan = August 2003, and blue = September 2003).
We find no systematic trend
with magnitude or color as shown in Figure \ref{RVcompplot}a--d.  
The bottom two plots in Figure \ref{RVcompplot} (panels e and f) show 
a comparison of $\Delta V_r$ vs.\ $V_r$.
While there may seem to be an odd trend at $-25 > V_r > -40$ km s$^{-1}$ in panel (e), 
this is due mainly to one cluster --- IC 4651 --- that has a peculiar offset.
This is demonstrated by the ``disappearance'' of that odd trend when IC 4651 is removed 
from the distribution (Figure \ref{RVcompplot}f; see \S \ref{comp2rv} for a detailed discussion 
of IC 4651).

Therefore, we find that our ``red'' sample, which we have made our standard reference,
 is consistent with previous work and this bears out our having corrected the blue sample RVs
 to the red RV system.
The cause of this offset it probably due to a combination of using ``blue'' 
standards from different runs, 
as well as the fact that the two blue Fekel standard stars have only a few good lines for RV determination combined with
large rotations with both Fekel stars having $V \sin i \sim 18$ km s$^{-1}$.


\section{Final Cluster Sample}
 
Table \ref{fullsample} summarizes all clusters observed, including UT date of the observation and exposure times, 
the numbers of stars selected to be cluster members by \citet[][]{dias01,dias02a} that were targeted
with Hydra fibers (col. 4),
the total number of stars and number of Tycho-2 stars observed (col. 5), 
the number of total observed stars with reliable RVs (col. 6), 
and the Tycho-2 stars with reliable RVs (col. 7).
For the WIYN data, we were able to observe nearly 75--80\% Tycho-2 stars used in the corresponding \citet{dias01,dias02a} 
survey.  
For the CTIO runs, we found that we were generally able to observe 50--80\% of the Dias et al.\ 
selected Tycho-2 stars, and, in addition, sample an average of $\sim 50$ more non-Tycho-2 stars 
(since the latter were generally 
fainter by 1--3 magnitudes, a lower fraction of them delivered reliable velocities in the allotted
observing time).
Since we are obtaining data for most of the Dias et al.\ stars, we will be able to compare our 
new membership data directly against the membership analysis done by these authors (see \S \ref{comp2dias}).


\section{Cluster Membership Analysis}

One of the most complicated problems affecting studies of open clusters is membership contamination 
associated with their location within the densely populated Galactic plane.  
Large numbers of disk stars unrelated to the cluster lie along the CMD sequences of the typical open cluster and, 
given the typical motions of many objects within the Galactic plane, usually with rather
similar velocities.  To determine the bulk motion of clusters one must first isolate true
cluster members from the dominant field star population in the fore/background.  
To accomplish this discrimination we have modified a previously implemented method designed to do 
just that.  The star's proper motion, RV, and spatial distribution are all used as inputs for a kernel-based,
probability distribution function
technique, described below, that eventually allows the cluster bulk motion to be determined 
from stars with high membership probabilities.  

\subsection{Non-Parametric Frequency Function}\label{datacuts}

To determine cluster membership probabilities for stars based on RV and proper motion, we have 
chosen to use an empirical, non-parametric technique --- modified from that described in \citet*{gjt98} --- 
that incorporates a kernel estimator \citep{hand82}
to isolate the phase space distribution of cluster stars in a field.  

\begin{figure} 
\epsscale{0.7}
\plotone{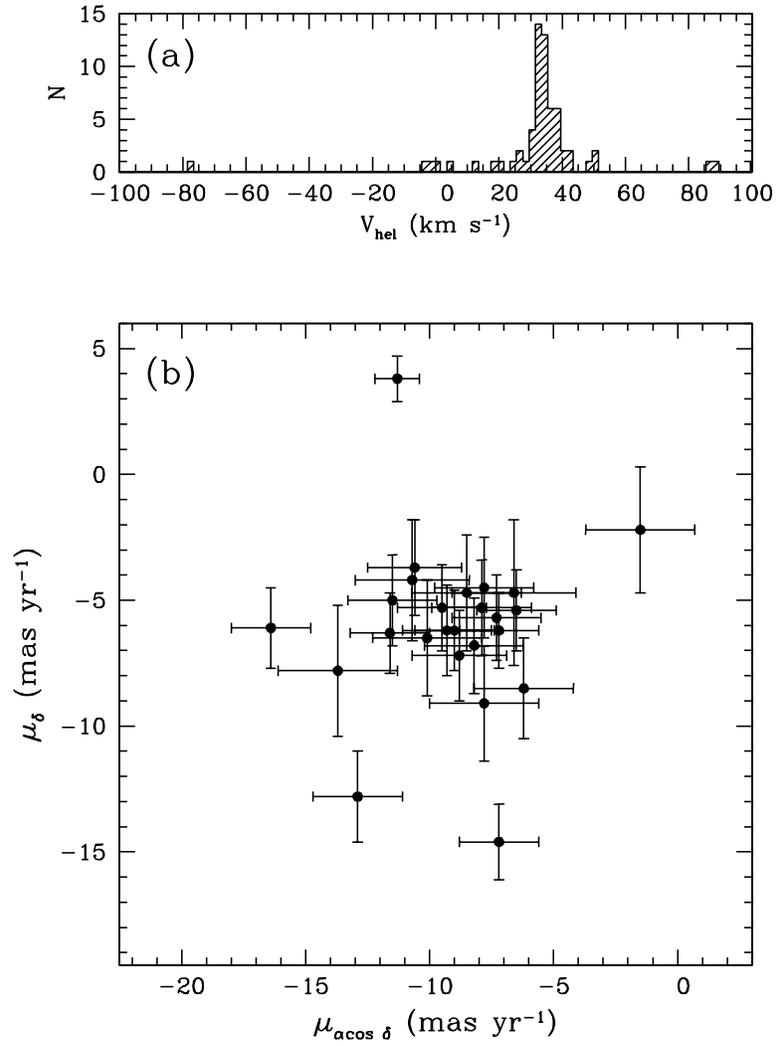}
\caption[Kinematic data for NGC 2682]{\label{rawM67} 
Our measured kinematical data for NGC 2682 (M67). 
(a) RV distribution, shown with 2 km s$^{-1}$ binning, 
of all stars with RVs between $-100$ and $+$100 km s$^{-1}$ 
measured for NGC 2682.
(b) Proper motion distribution of all observed stars with proper motion data in the NGC 2682 field 
having Tycho-2 data with error bars shown.}
\end{figure}

While we adopt the basic technique used by \citet{gjt98} for proper motions alone, 
we have generalized it also to operate on a spatially-constrained, 1-D RV distribution as well as an 
RV-constrained, 2-D proper motion distribution.
In principle, one could use either distribution separately for culling cluster members, 
but for the most secure assessment of membership we depend on the joint probability distributions. 
This means, therefore, that we can only use stars
having both RV and proper motion data.  To improve our results further, 
we remove stars with large measurement errors in either proper motion or RV, or those stars that clearly 
have halo-like RVs, 
using the following constraints applied to the data:
\begin{itemize}
\item $\mu$ error limit: $\sqrt{\sigma^2_{\mu_{\alpha}^*} + \sigma^2_{\mu_\delta}} \le 10$ mas yr$^{-1}$
\item RV limit: $-200 < V_r < 200$ km s$^{-1}$
\item RV error limit: $\sigma_{V_r} \le 10$ km s$^{-1}$
\end{itemize}
The modified version of the \citet{gjt98} formulation is intended to
 perform better for our particular survey circumstances --- 
i.e.,  fewer numbers of stars per cluster, 
but high-quality RV data for these stars. 
Throughout the following description we will demonstrate the 
basic features of our analysis via the example processing of the cluster
NGC 2682 (M67), for
which the raw data are shown in Figure \ref{rawM67}.

\subsection{1-D Kinematical Distribution: Radial Velocities }

For our data, the RV distribution is found to be the most sensitive discriminator of cluster
membership because of the small measured relative RV errors.
When applying a kernel density estimator the empirical density function ($\psi^V_{c+f}$) is comprised of both 
the cluster ($c$) and the field ($f$), where here $V$ stands for the RV distribution.  
Since the observed empirical density function is the sum of two underlying distributions 
(e.g., $\psi^V_{c+f}  = \psi^V_{f} + \psi^V_{c}$), one must decompose the    
distributions to isolate the cluster function.  
Because of the accuracy of the RV data and the small intrinsic velocity 
dispersions of open clusters (0.5--3 km s$^{-1}$),
we expect to be able to discriminate the cluster and field fairly readily.  
To do so, however we must first isolate the field population 
to verify which peak in the $\psi^V_{c+f}$ distribution is due to the cluster.
Differences in the cluster versus field distribution should be evident by looking 
at samples of stars drawn from different radii from the cluster center. 
A useful initial assumption is that stars outside of the cluster radius are ``non-members'', 
and these can provide a reasonable estimate of $\psi^V_{f}$.

\begin{figure} \epsscale{1.0}
\plotone{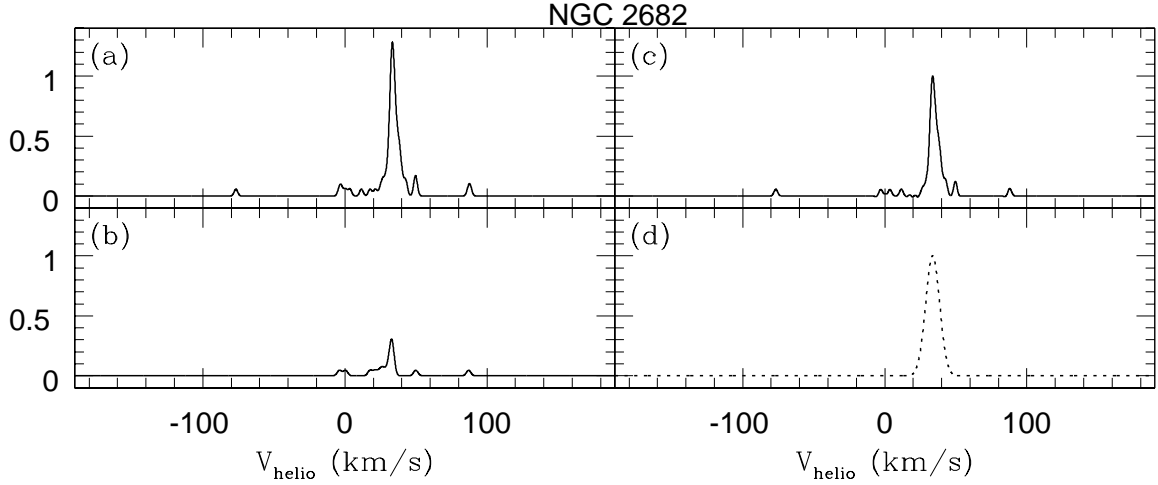}
\caption[Kernel output from 1-D RV distribution for NGC 2682]{\label{rvM67} 
Steps in the membership analysis for NGC 2682 (M67) for the 1-D RV distribution ($V_r$).
(a) Kernel-smoothed RV distribution for all stars used in the analysis of the data for the cluster NGC 2682.
(b) Kernel-smoothed distribution for stars not within the cluster radius \citep{dias02b}. 
(c) Probability distribution estimated by [(a)$-$(b)]/(a).
(d) 1-D Gaussian fit to (c).  The fit to the ``cluster'' distribution used to determine the
membership probability ($P_c^V$) based on the spatially-constrained RV data. 
}
\end{figure}

The RV data kernel analysis is comprised of four steps: 
(1) All RV data are convolved with a Gaussian kernel to homogenize our errors for a given 
cluster.   This kernel has a width 
determined by the mean RV errors from all of the observed stars in a given cluster field.
Because open clusters have intrinsic velocity dispersions of 1--3 km s$^{-1}$
in addition to our measurement errors, we limit 
the Gaussian width to be {\it at least} 3 km s$^{-1}$ and at most 10 km s$^{-1}$.
Applying the Gaussian kernel to smooth our RV data ($\psi^V_{c+f}$) produces the 
smoothed field plus cluster distribution $\Psi^V_{c+f}$; 
an example for NGC 2682 (M67) is shown in Figure \ref{rvM67}a. 
(2) We apply the same Gaussian kernel to smooth the RV data of stars that are outside the cluster radius 
\citep[utilizing the cluster diameters from ][]{dias02b}.  This smoothed RV distribution is used as 
the field distribution $\Psi^V_{f}$(Figure \ref{rvM67}b). 
(3) We wish to determine the probability of any particular star with a given RV 
being a member of the cluster, so we need to determine the 
normalized probability distribution: 
\begin{equation}
P^V_c (V_{r,i}) = \frac{\Psi^V_{c+f} (V_{r,i}) - \Psi^V_{f} (V_{r,i})}{\Psi^V_{c+f} (V_{r,i})}.
\end{equation}
The cluster probability distribution $P^V_{c}$  is shown Figure \ref{rvM67}c); 
however, we see that a few outliers, which are non-member stars within the cluster radius, 
are still visible in the distribution.  
(4) We assume that the strongest peak in the ``cluster'' probability distribution, $P^V_c$, 
belongs to the cluster,  
and perform a 1-D Gaussian fit to this peak  
(Figure \ref{rvM67}d; dotted line).  
This Gaussian fit is used to determine RV membership probabilities, $P_c^V$, for all stars in the field and 
to exclude non-member RVs that still may appear in $\Psi^V_{c}$.

\subsection{2-D Kinematical Plane: Proper Motions}

\begin{figure} \epsscale{1.1}
\plottwo{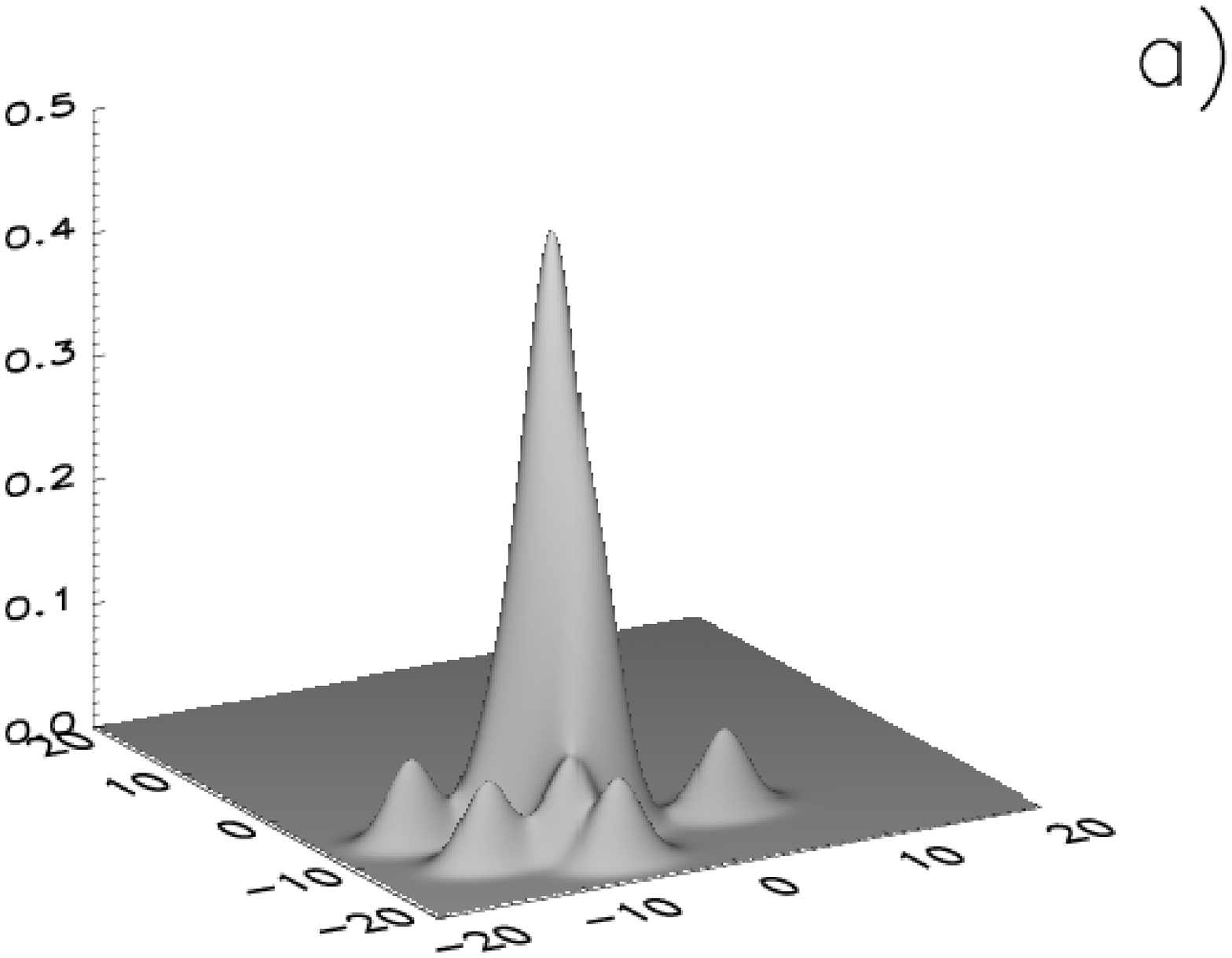}{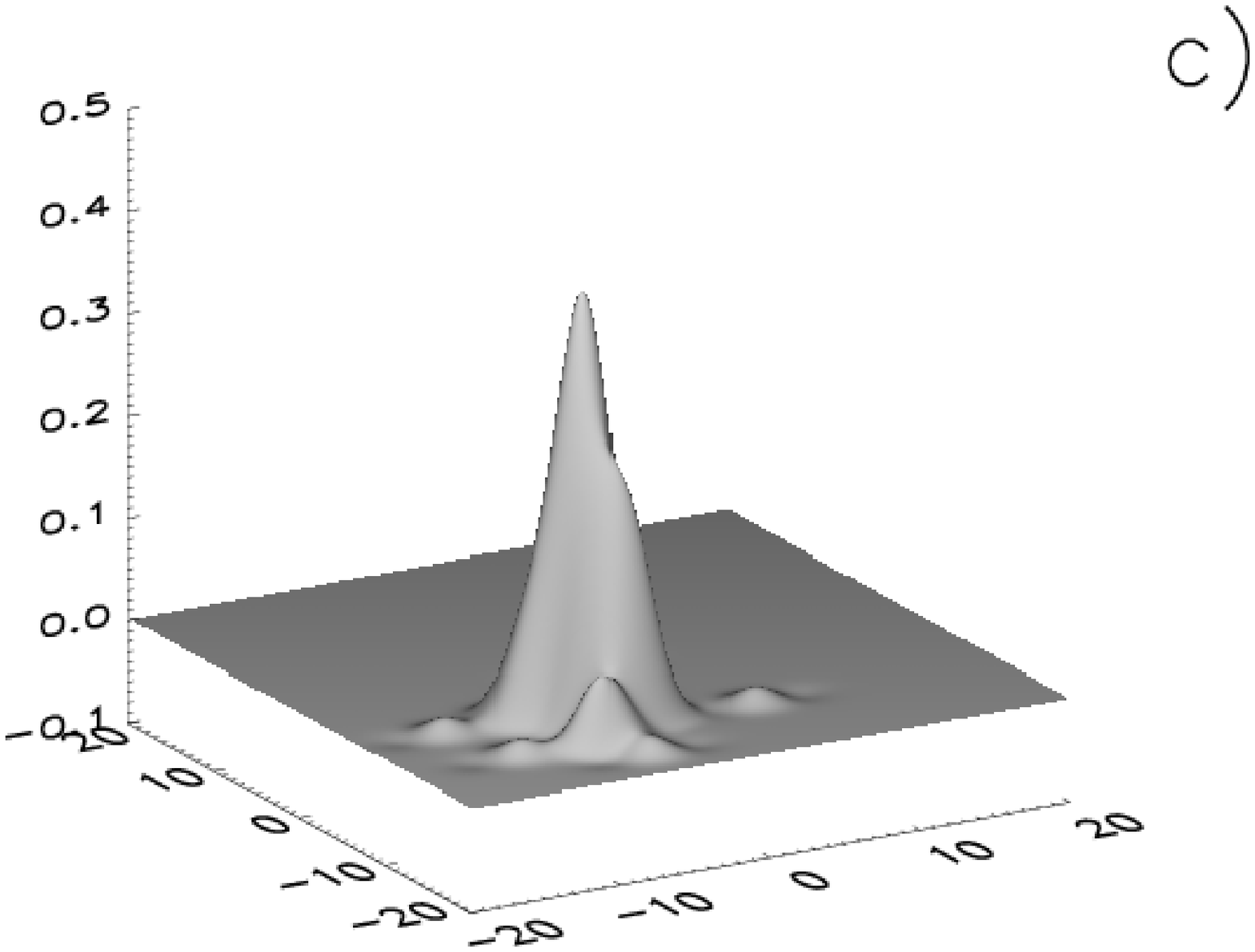}
\plottwo{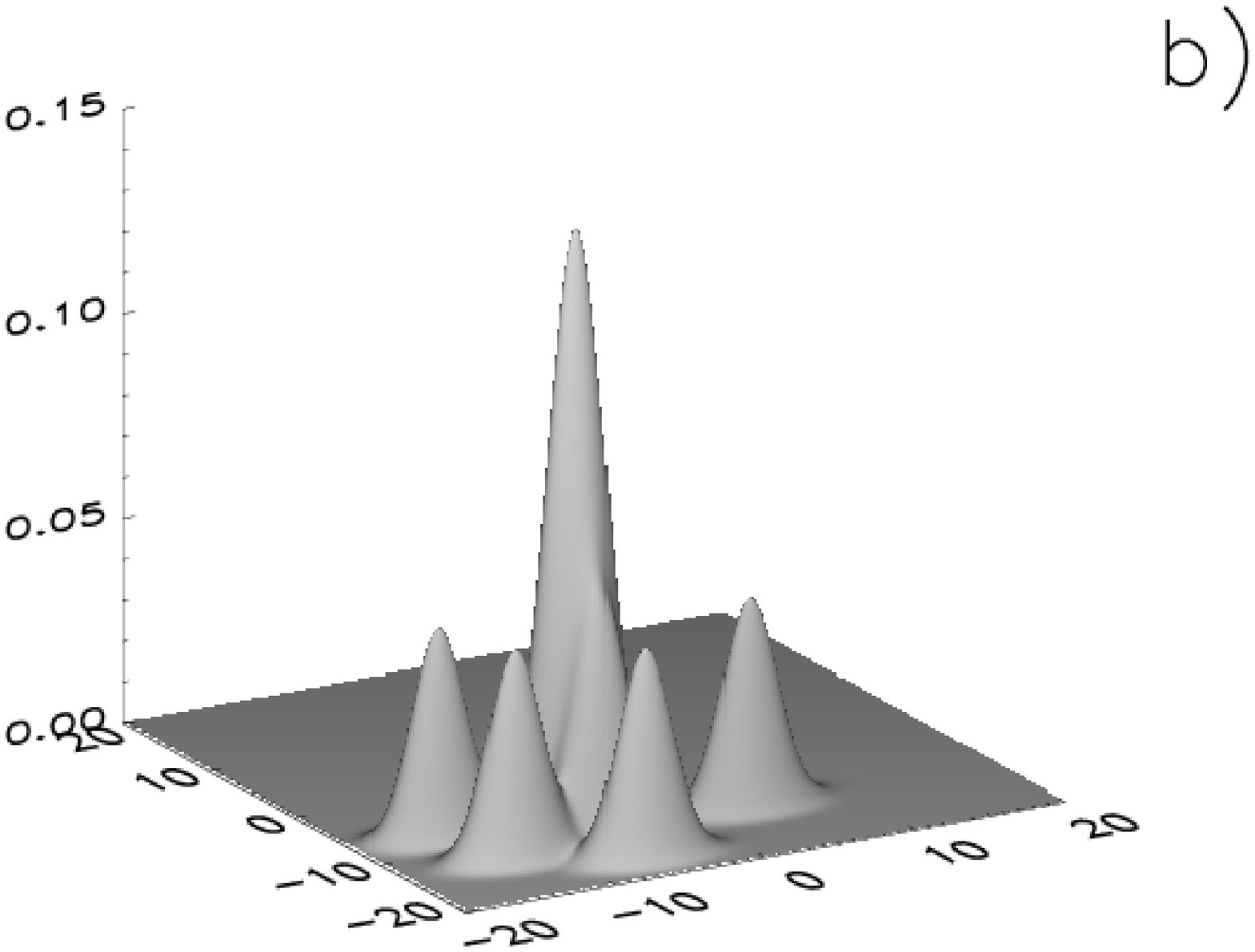}{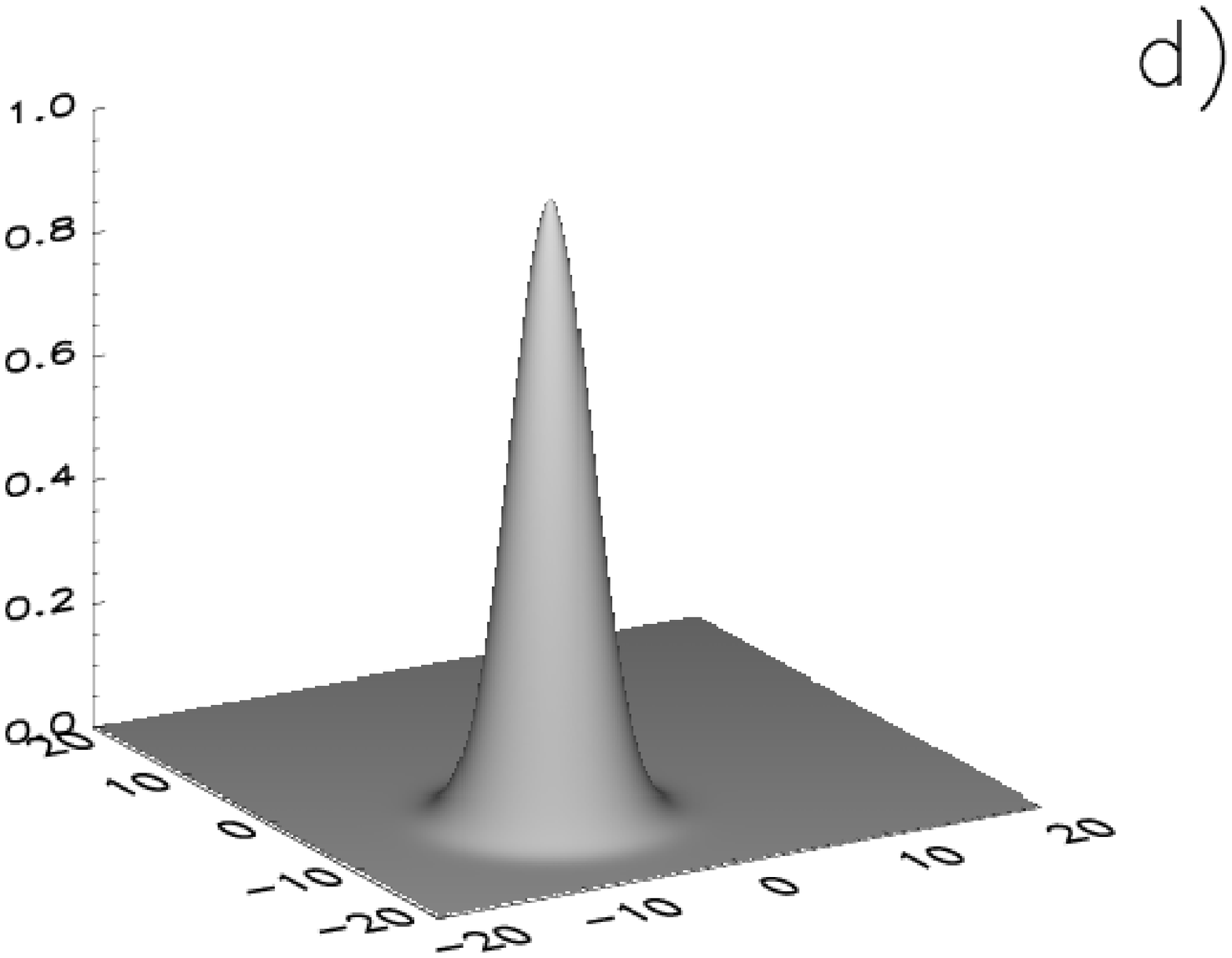}
\caption[Kernel output from 2-D PM space for NGC 2682]{\label{pmM67} 
Steps in the analysis of the 2-D proper motion distribution ($\mu_{\alpha \cos(\delta)}$, $\mu_{\delta}$) 
for NGC 2682 (M67). 
(a) Kernel-smoothed distribution of all stars used in the analysis.
(b) Kernel-smoothed distribution for stars not selected to be RV members (see Figure \ref{rvM67}; $P_c^V < 0.8$). 
(c) The probability distribution given by [(a)$-$(b)]/(a).
(d) The normalized 2-D Gaussian fit to the distribution in panel (c).  The resulting fitted ``cluster'' distribution used to determine $P_c^K$.}
\end{figure}

The proper motion kernel analysis is similarly comprised of four steps, but now applied in 2-D.
This technique for proper motions is identical to that used in \citet{gjt98} with the exception that 
instead of using a spatial membership separation (which we used for the RV distribution described above) 
to establish $\Psi^K_{f}$, 
we have chosen to use the RV separation described above (i.e., those stars outside the 
Gaussian fit to the RV distribution, $P_c^V = 0$, are considered the ``field'' population). 
The proper motion kernel uses the following equation analogous to that used in the RV analysis above:
\begin{equation}
P^K_c (\mu_{\alpha' , i},\mu_{\delta, j}) =
 \frac{\Psi^K_{c+f} (\mu_{\alpha', i},\mu_{\delta, j}) - \Psi^K_{f} (\mu_{\alpha', i},\mu_{\delta, j})}{\Psi^K_{c+f} (\mu_{\alpha', i},\mu_{\delta, j})},
\end{equation}
where $\alpha'$ is $\alpha \cos(\delta)$.
Continuing our example of NGC 2682, we apply the 2-D kernel smoothing to the proper motion distribution 
as shown in Figure \ref{pmM67}a-d.
The Gaussian fit to the field-subtracted distribution is used to determine proper motion membership 
probabilities $P_c^K$ (Figure \ref{pmM67}d) for stars in each cluster. 
Both the RV and  proper motion kernel analysis was performed on all clusters in Table 6.

\subsection{Calibration of the Membership Criteria}

To determine the RV membership ``cutoff'' criteria, we have chosen to analyze in detail one of our best sampled clusters, 
our example NGC 2682.
Using techniques standard dynamical techiques from \citet{pm93}, we performed an iterative 3$\sigma$ rejection using the 
full sample of NGC 2682 RV data.  We find an intrinsic velocity dispersion of $\sigma_{int} = 0.96 \pm 0.29$ km s$^{-1}$.
As a comparison, using proper motions, \citet[][]{girard89} found that NGC 2682 (M67) has $\sigma_{int} = 0.81 \pm 0.10$ km s$^{-1}$.
Comparing the RV member stars left after applying the iterative 3$\sigma$ rejection, we find that all of the remaining stars have $P_c^V \ge 70$\%.  
More lenience is given to the proper motions due 
to the larger average error, and in this regard
we follow the criterion used by \citet{dias01,dias02a}.  
As a result, we have chosen to 
define cluster membership as stars that have $P_c^V > 70\%$ and $P_c^K > 51\%$.

\subsection{Results of the Membership Analysis}
Cluster membership was determined by jointly assessing the probabilities from the 1-D RV distribution ($P_c^V$) 
and the 2-D proper motion distribution ($P_c^K$).
The probabilities for each star analyzed in the cluster NGC 2682 are included along with the
RV and proper motion data in Table \ref{table:NGC_2682}.
This table includes the star name from the Tycho-2 survey, 
or if not a Tycho-2 star, another identifier \citep*[for M67 we have IDs from][]{es64,sand77,mmj93,fbc}. 
The table then
 lists, in order, the right ascension and declination for each M67 star
 (cols. 2 and 3), the Tycho-2 
proper motions and errors (cols. 4--7), our measured RV and error (cols. 8 and 9), 
and which spectral cross-correlation 
template was used to derive these (col. 10).
In addition, we have included the membership probability from Dias et al.\ (2001, 2002a; col. 11) 
for comparison to our derived membership probabilities 
$P_c^K \times 100$ (PM; col. 12), $P_c^V\times 100$ (RV; col. 13), and $P_c^{tot}$, the
joint probability ($P_c^{tot} = P_c^V P_c^K \times 100$; col. 14).
The stars selected as cluster members are presented in boldface type.
 
Similar probability data are given for the other clusters in our sample
in Table \ref{alldata}, which is available in electronic format.
In this table we give for each star observed its Tycho-2 name, or, if a non-Tycho star,
an identifier with the format 
``XXXX\_f\_\#\#\#\#''  for the added ``filler'' candidate FOPS guide stars or ``XXXX\_u\_\#\#\#\#'', for 
USNO B-1.0 catalog ``filler'' stars, categories described in the observational criteria in \S \ref{targselect}.
In \S \ref{comp2dias} our analysis of the cluster memberships of these stars are compared against 
the membership analysis by \citet{dias01}, 
whose membership probabilities are based only on proper motion.

\subsection{Cluster Membership and Cluster CMDs}

As shown in Figure 1, with only photometric data 
the identification of open cluster sequences in the CMD can
often be a tricky prospect.  Our radial velocity cluster memberships
can significantly aid in clarifying the location of these cluster sequences.
The 2MASS and Tycho-2 photometry for all stars in our survey with measured RVs 
are listed in Table \ref{ALLPHOT}.   

\begin{figure} %
\epsscale{1.0}
\plotone{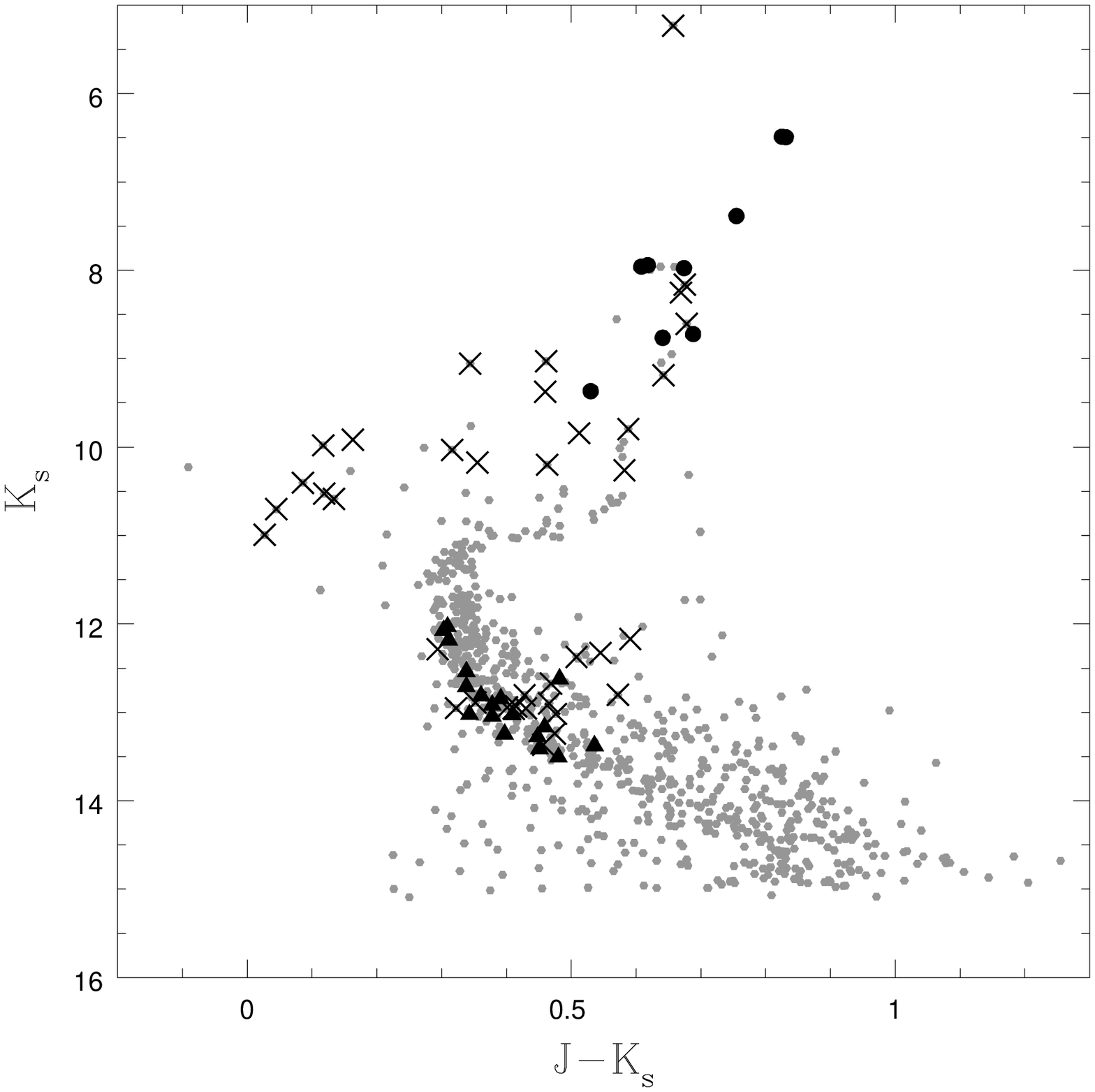}
\caption[NGC 2682 Membership Selection and 2MASS CMD]{\label{M67plot}  
2MASS color-magnitude diagram (CMD) for NGC 2682 (M67) for stars 
inside the cluster radius \citep{dias02b}. 
Crosses ($\times$) denote stars that we determined to be non-members. 
Large circles denote stars selected to be members based on {\it both} RV and proper motion criteria.  
Triangles denote stars that have $P_c^V > 70$\% but which do not have Tycho-2 proper motion data available.
}
\end{figure}

\begin{figure} %
\epsscale{0.99}
\plotone{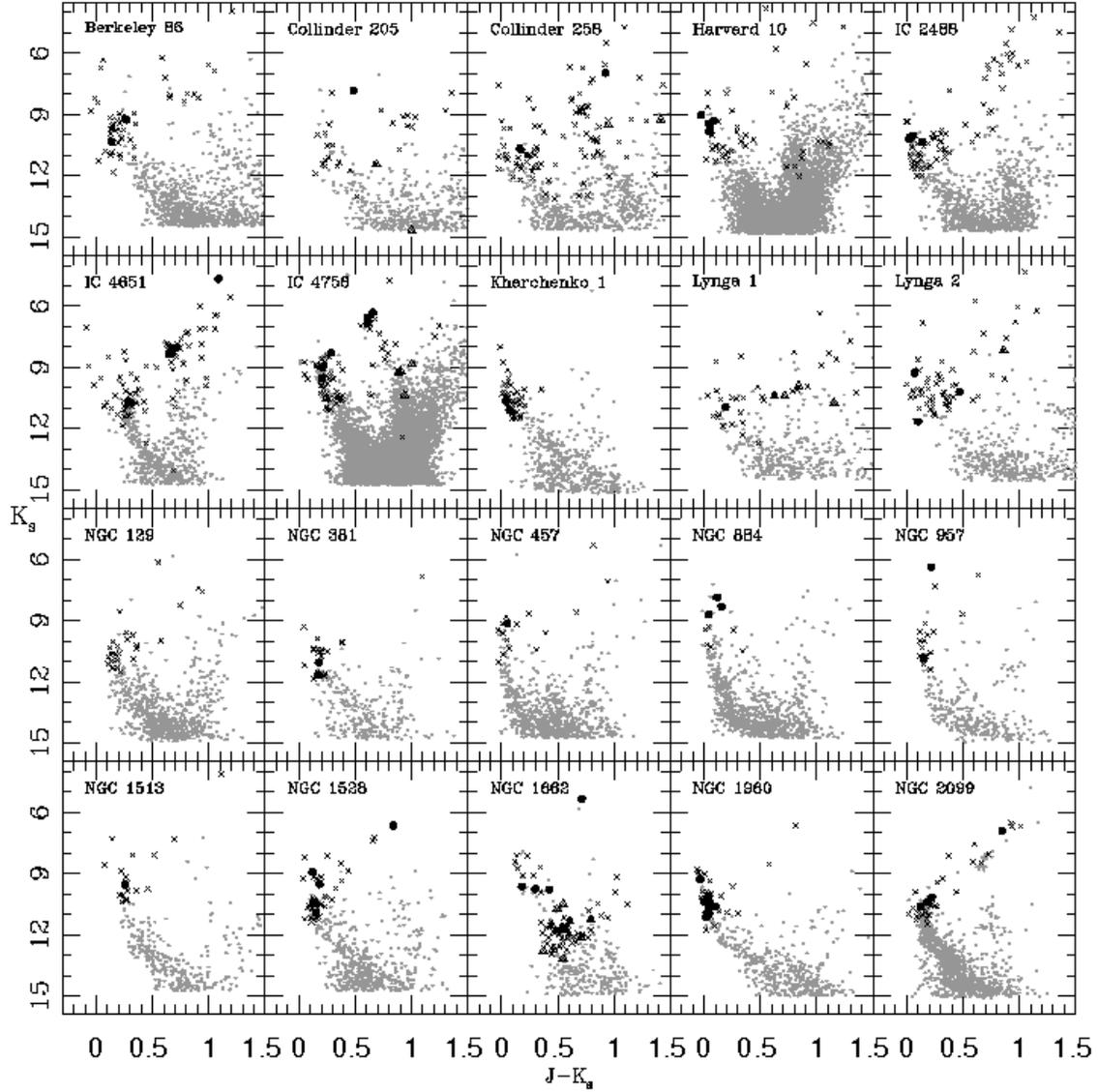}
\caption[Membership Selection and 2MASS CMD]{\label{allCMD1}  
2MASS color-magnitude diagram (CMD) for all clusters using stars 
inside the cluster radius \citep{dias02b}. 
Crosses ($\times$) denote stars with proper motion data that we determined to be non-members. 
Large circles denote stars selected to be members based on {\it both} RV and proper motion criteria.  
Triangles denote stars that have $P_c^V > 70$\% but which do not have Tycho-2 proper motion data available.
}
\end{figure}

\begin{figure} %
\epsscale{0.99}
\plotone{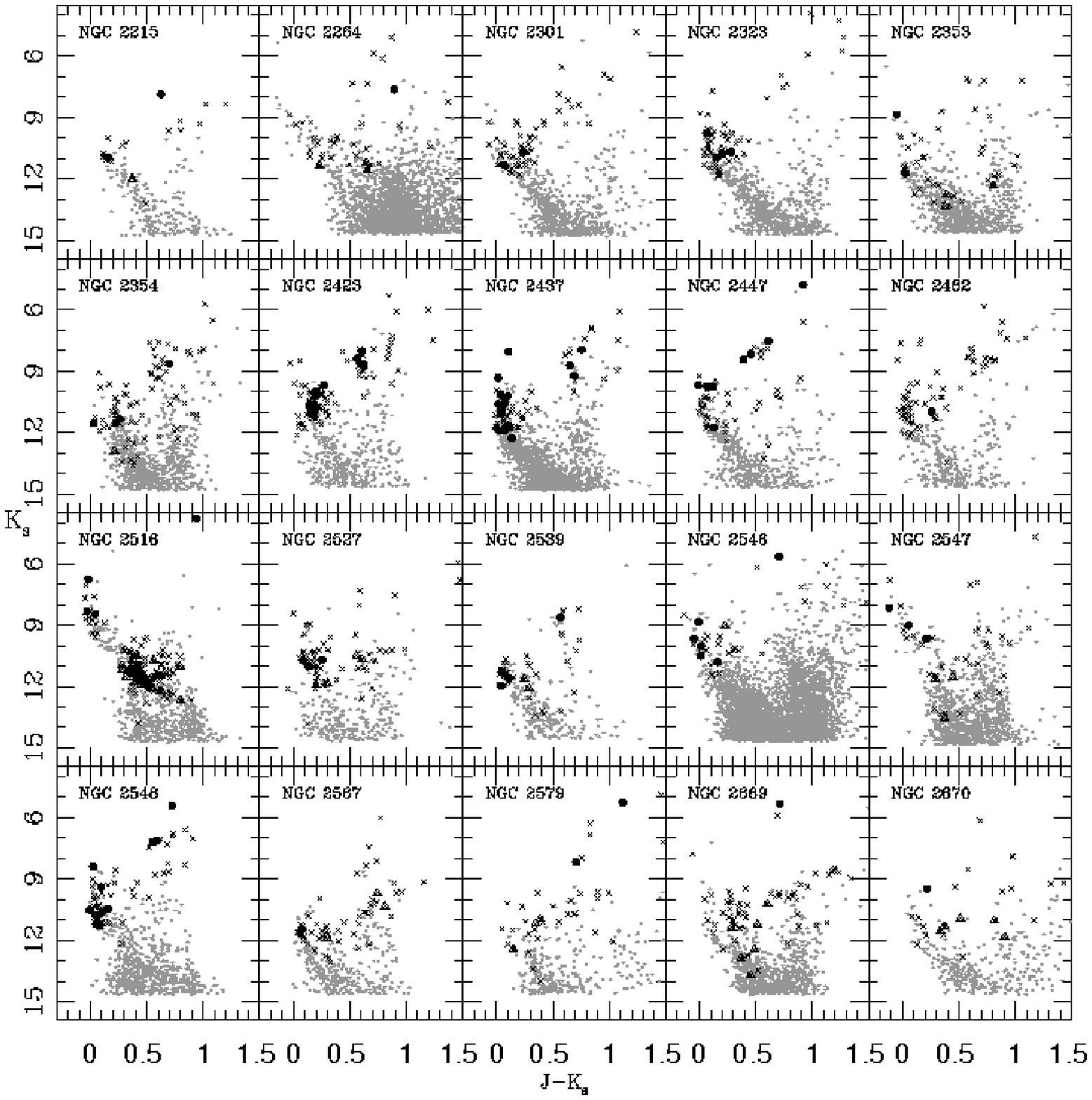}
\caption[NGC 2682 Membership Selection and 2MASS CMD]{
Same as Figure \ref{allCMD1}.\label{allCMD2}  }
\end{figure}

\begin{figure} %
\epsscale{0.99}
\plotone{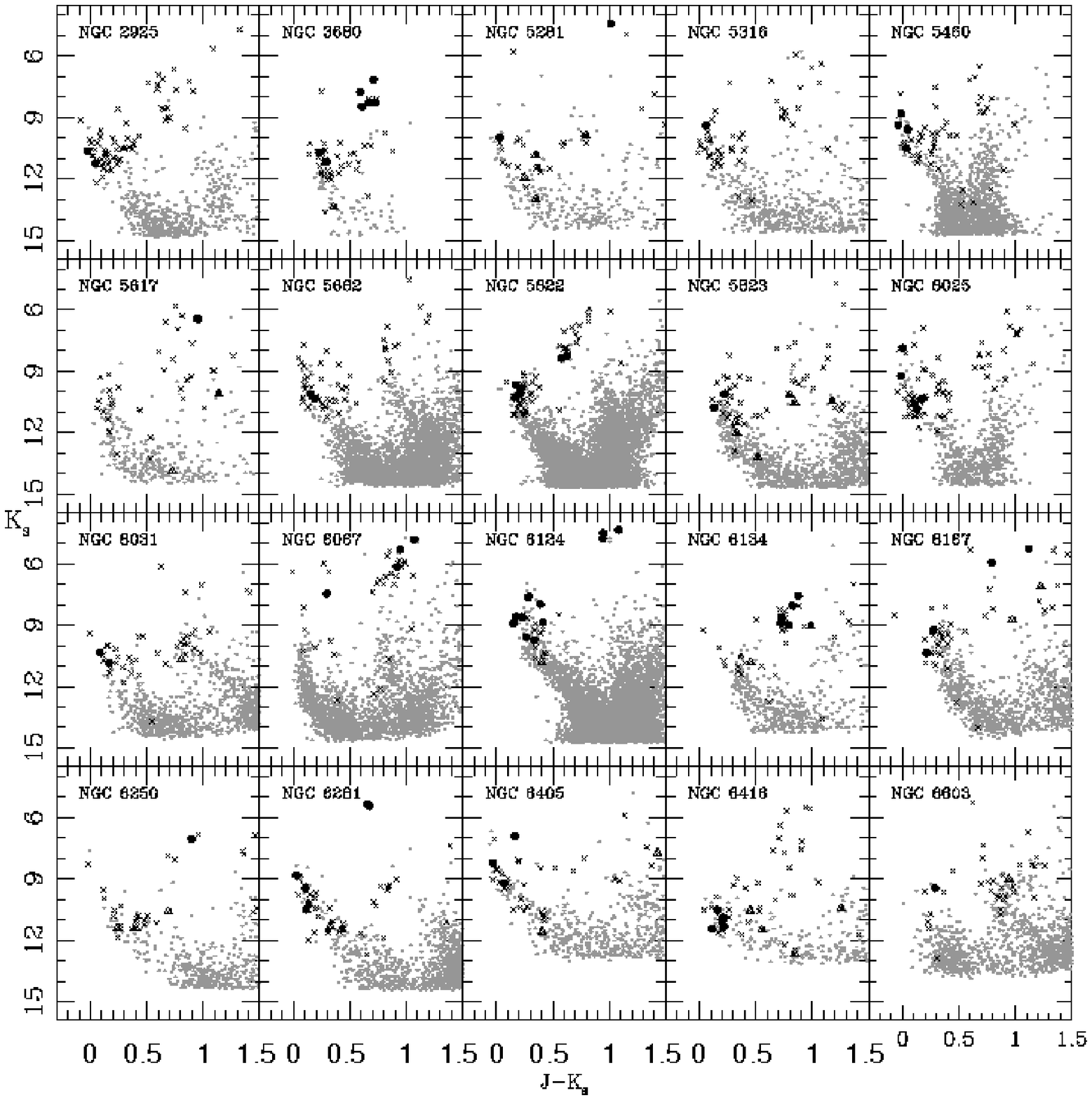}
\caption[NGC 2682 Membership Selection and 2MASS CMD]{
Same as Figure \ref{allCMD1}.\label{allCMD3}  }
\end{figure}

\begin{figure} %
\epsscale{0.99}
\plotone{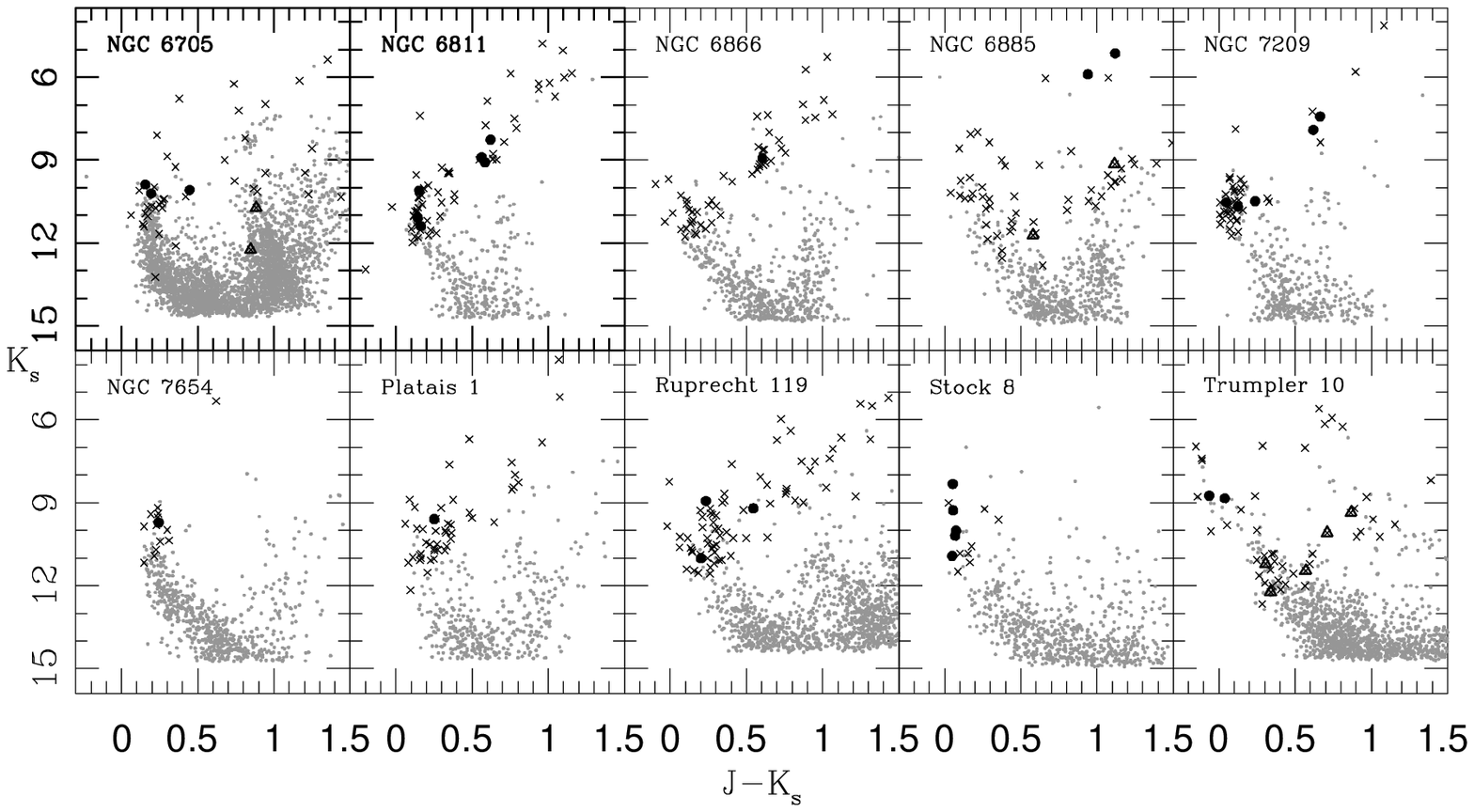}
\caption[NGC 2682 Membership Selection and 2MASS CMD]{
Same as Figure \ref{allCMD1}.\label{allCMD4}  }
\end{figure}

Figure \ref{M67plot} shows the 2MASS CMD for the example cluster NGC 2682
with our spectroscopically-observed stars identified, and with
large circles denoting stars selected to be members based on {\it both} RV and proper motion.  
Triangles denote stars that have $P_c^V \ge 70$\% but which do not have Tycho-2 proper motion data.  
For now we present CMDs without reddening corrections applied, because this is a non-trivial process
in that not all line-of-sight reddening \citep*[the values typically given in catalogs such as][]{schlegel98} is necessarily foreground to the cluster.
One can see from the CMD 
that in this case our membership census yields members 
that fall primarily along the photometric sequences of M67 apparent in the CMD.
Similar 2MASS CMD membership plots for all clusters we have studied are shown in
Figures \ref{allCMD1}--\ref{allCMD4}. 
 As in the case of M67, our
identified members typically fall in the expected locations of the main sequence turn-off (MSTO) or
giant branches of the clusters, when those are obvious; however, in many cases the CMDs
are crowded with field star contamination and our identified members help clarify the cluster
sequences.  This is particularly useful in the fairly common situation where the giant branches
are sparsely populated.  As we shall show in another contribution (Frinchaboy et al., {\it in prep}), our ability to 
clarify the CMD locations of cluster giant branches and MSTOs greatly improves the 
isochrone fitting for these systems.

\section{Kinematical Results}

\subsection{Derived Cluster Space Velocities}\label{spacemotions}
 
The cluster bulk RV is calculated using cluster members (e.g., as shown in Table \ref{table:NGC_2682}) and techniques from \citet{pm93} 
to determine the cluster mean RV and error in the mean.  
The cluster mean bulk proper motions are calculated using the following equations 
(and a symmetrical version for $\mu_{\delta}$).
\begin{equation}
< \mu_{\alpha \cos(\delta)} >  = \frac{\sum_{i=1}^{n} \left(\frac{\mu_{i,\alpha \cos(\delta)}}{\sigma_{\mu_{\alpha \cos(\delta)},i}^{2}}\right)}{ \sum_{i=1}^{n} \left(\frac{1}{\sigma_{\mu_{\alpha \cos(\delta)},i}^{2}}\right)} 
\end{equation}
\begin{equation}
\epsilon_{\mu_{\alpha \cos(\delta)}} = \frac{1}{ \sum_{i=1}^{n} \left(\frac{1}{\sigma_{\mu_{\alpha \cos(\delta)},i}^{2}}\right)}.
\end{equation}

The derived cluster bulk motions are given in Table \ref{bulkALL}, where we 
list the numbers of members with full space motions (col. 2) and the 3D members plus the 
stars determined to be members by RV criteria alone (3D+RV; col. 3),  
along with the resulting bulk kinematics 
and the associated uncertainties (RV from all 3D members; col.\ 4), 
RV from 3D {\it and} additional ``RV only'' members (col.\ 5), 
and equatorial and Galactic system proper motions (cols.\ 6--9). 
We find two clusters NGC 1513 and NGC 7654 with only one star selected for membership (i.e., the membership method 
found no more than one star with a given RV within the errors);
given the uncertainty in selecting among single star subsamples to define 
the actual ``cluster'', we remove these two clusters from further analysis.

\subsection{Comparison to Previous Results}

\subsubsection{NGC2682 (M67) Example}

In \S3.7 and Table \ref{clM67comp} we have already demonstrated a star-by-star comparison
of derived RVs for the example cluster M67.  
For stars in common between the surveys, we find excellent agreement in the determined per star RVs 
(previously shown in Figure \ref{RVcompplot}).  
Now we compare the derived bulk space velocity for this very well-studied cluster to the most
detailed, previous studies of M67.

In Table \ref{clcomp2} we compare our derived mean proper motion and radial velocity for M67, 
averaged over these measured parameters for 10 stars we determined to be reliable 3D members
of the cluster, 
against derivations of these bulk motion parameters by other authors.
With regard to to the previously derived bulk RV for M67, 
our mean radial velocity is consistent
with previous measurements by \citet{mathieu86} and \citet{sfj95}, and lies
within 0.2 km s$^{-1}$ of the rather precise value given in the 
Mathieu et al.\ study.  
The total number of published clusters having as
 extensive and detailed RV coverage as the Mathieu et al. M67 study is less than ten, whereas
our study now provides 
high precision RVs for stars in nearly five times as many clusters. 
We also find proper motion results more or less consistent with previous measured values, with our
$\mu_{\delta}$ value being bracketed by the $\mu_{\delta}$ measurements by
 \citet{dias01} and \citet{khar05} results and our $\mu_{\alpha} \cos{\delta}$ reasonably close to the
 values for this proper motion component derived by these two other studies.
The previous studies have smaller errors 
in their mean due to the larger numbers of ``member'' stars used in the 
determination of the bulk proper motion.

Thus we find that our survey results are consistent with the very detailed analysis of previous M67 work.
Despite the fact that M67 is probably one of the most well-studied open clusters in the Galaxy and  
previous studies typically utilized many more stars than we have, our results deliver comparable
precision to the best of these because of the greater purity of our samples, and, in the case of
the RV measurement, the velocity resolution of our spectra.

\subsubsection{Comparison to Previously Derived Bulk Cluster Radial Velocities}\label{comp2rv}

\begin{figure}\epsscale{1.01}
\plotone{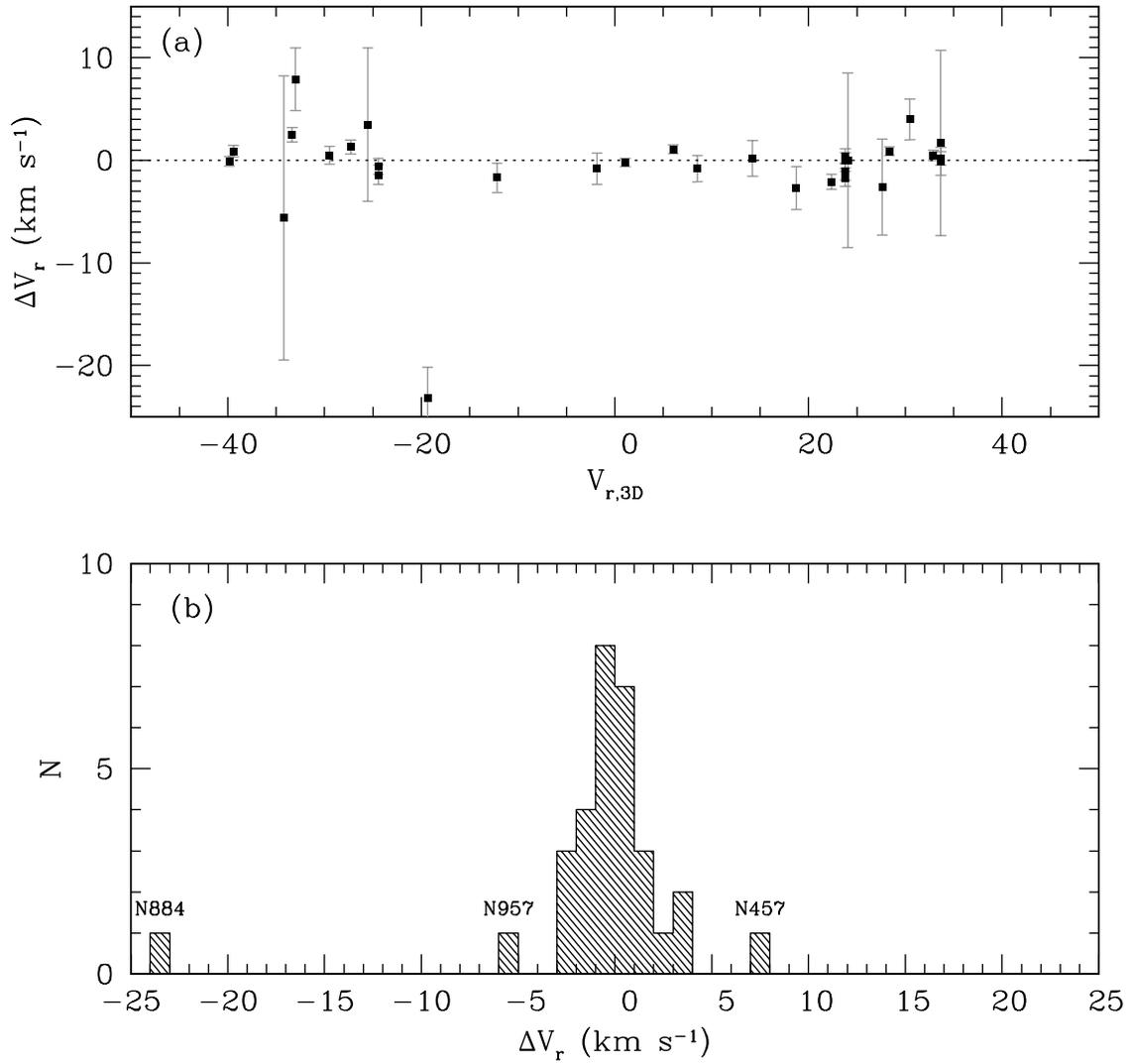}
\caption[Comparison of final cluster radial velocities.]{\label{RVvRVFinal}Comparison 
of radial velocities $\Delta V_r$ to previous studies (see Table \ref{bulkCompare}).   
(a) $\Delta V_r$ plotted as a function of our measured $V_r$; no obvious systematic trend is seen.
(b) Histogram of  $\Delta V_r$ showing that, besides the cases of NGC 457, NGC 884, and NGC 957 
which all compared here to the results by \citet[][see \S \ref{comp2rv}]{li89},
all of our measurements of the bulk RVs of the clusters are within 5 km s$^{-1}$ of 
all previous determined cluster measurements, and with the peak at $\Delta V_r = 0$ km s$^{-1}$.
}
\end{figure}

A compilation of our derived mean cluster RVs compared to those found previously by other
authors is given in Table \ref{bulkCompare}.  We have found previous results for 25 of our 71
studied clusters, some with multiple studies.  In general, we find consistency with the previous
studies to the few km s$^{-1}$ level as shown in Figure \ref{RVvRVFinal}. 
but in a few cases, there are more substantial differences.

Figure \ref{RVvRVFinal} shows that
 the clusters NGC 457, NGC 884, and NGC 957 have discrepant RVs found between our work and any previous study; however  
all of these clusters, plus NGC 2264, were studied by \citet*{li89}.
The \citet{li89} study is comprised of only a few possible cluster members observed 
(e.g., for NGC 884 and NGC 957 only two stars each and these clusters also have large mean errors).
We believe our results, which incorporate both RV and proper motion membership, are superior to 
those from \citet{li89}.
Even with the small numbers of stars in both studies, 
we find that our results are marginally consistent with \citet{li89} for 
NGC 2264.

\subsubsection{Comparison to Previously Derived Bulk Cluster Proper Motions}\label{comp2dias}

\begin{figure}\epsscale{1.01}
\plotone{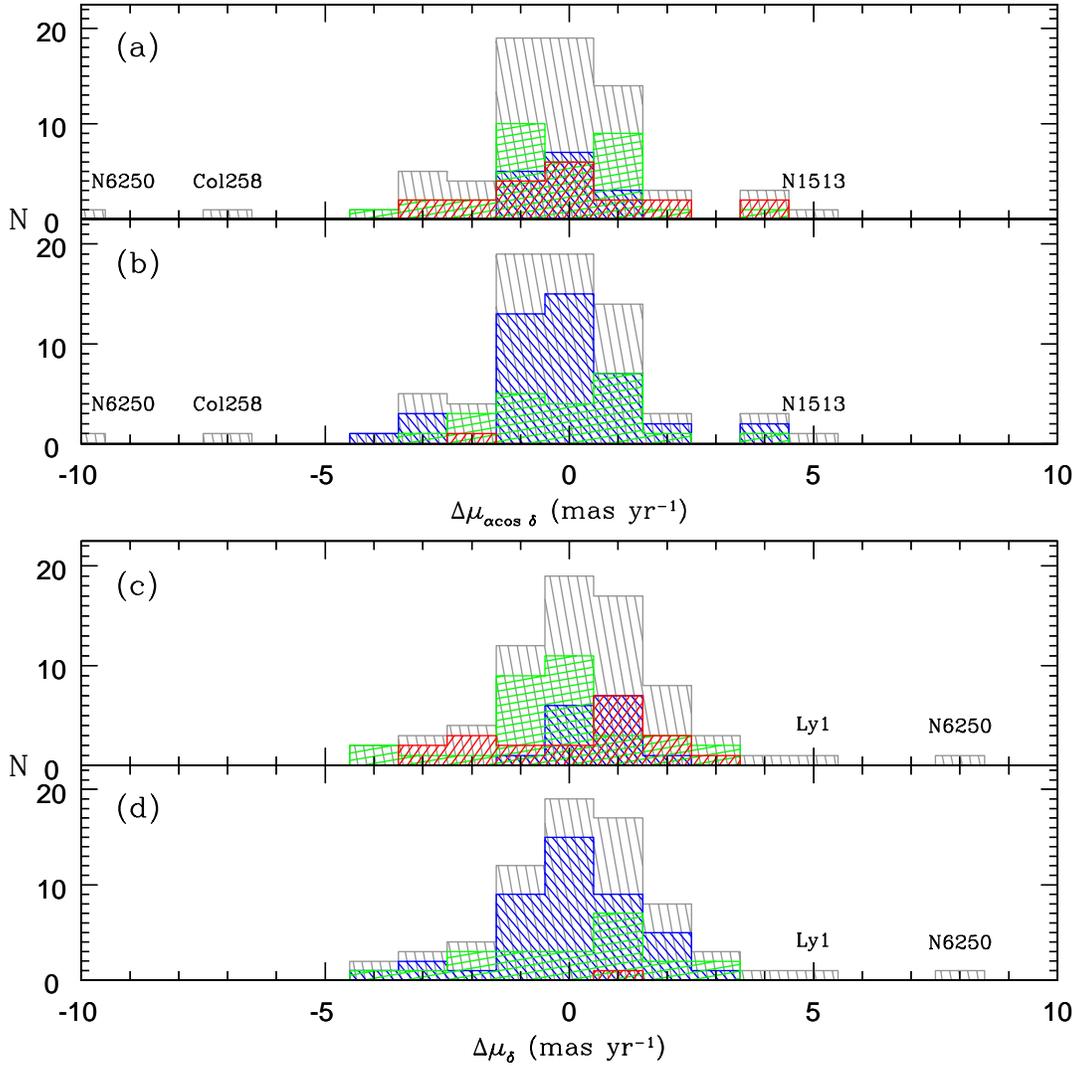}
\caption{\label{PMvPMHist} Comparison of the proper motions 
$\Delta \mu_{\alpha \cos \delta}$ and $\Delta \mu_{\delta}$ derived from our study to those
of \citet[][Table \ref{diascomp}]{dias01,dias02a}.
Histograms of  $\Delta \mu$ showing that, besides the cases of Collinder 258, Lynga 1, and NGC 6250,
all of our reliable measurements of the bulk RVs of the clusters are within 5 mas yr$^{-1}$ of 
\citet[][]{dias01,dias02a} study with the peak at $\Delta \mu_{\alpha} = 0$ mas yr$^{-1}$ and $\Delta \mu_{\delta} = 0$ mas yr$^{-1}$.
(a) $\Delta \mu_{\alpha}$ showing clusters having seven or 
more 3D members from our own analysis (blue histogram), while the green histogram denotes clusters with
3--6 3D members, and red histogram showing those clusters with less than three 3D members. 
(b) as (a) with histograms color-coded by membership from Dias et al., with the blue histgram having $\ge 20$ members, 
green 10--19 members, and red $< 10$ members.
(c) $\Delta \mu_{\delta}$ with same color-coding as (a).
(d) $\Delta \mu_{\delta}$ with same color-coding as (b).
}
\end{figure}

In Table \ref{diascomp}, we compare our derived open cluster 
bulk proper motions with the previous results of \citet{dias01,dias02a}.  The latter surveys
used only the Tycho-2 proper motions to derive membership and the cluster bulk proper motions.
Table \ref{diascomp} compares the 
numbers of stars used by Dias et al.\ and their derived mean cluster proper motions 
(col. 2--5) to our own sample statistics and derived mean proper motions (col. 6--9).
As shown in Figure \ref{PMvPMHist} (grey histogram), three clusters --- Collinder 258, Lynga 1, and NGC 6250 ---
show large inconsistencies ($\Delta \mu > 5$ mas yr$^{-1}$) between our results and those of Dias et al.
We also reminder the reader that we have already excluded two other cases (NGC 1513 and NGC 7654; see \S \ref{spacemotions}) from
our study, because we identified only one star selected as a possible cluster member.
Looking further at the proper motion difference outliers, 
we find that each Lynga 1 and NGC 6250 have only one star
with fully derived 3D kinematics and in the case of Collinder 258 there are only two member stars.
Thus, we conclude that our analysis may have settled on the wrong star(s) 
to represent the cluster in these cases
and that the results for Collinder 258, Lynga 1, and NGC 6250 (in addition to NGC 1513 and NGC 7654) 
may not be reliable.
For the remaining 66 of our 71 clusters, our ``re-measured'' proper 
motions are within the $1 \sigma$ errors of those found by \citet{dias01,dias02a}, though our data generally 
have comparable or smaller resulting errors in the mean (as shown in Figure \ref{PMvPMHist})
of $\sim 1.5$ mas yr$^{-1}$.

The direct comparison to the Dias et al.\ proper motions is shown in Figure \ref{PMvPMHist}, 
with the full sample shown in grey and various subsamples based on the number of members in either
survey shown by the colored histograms.
A somewhat close agreement with Dias et al.\ is expected because we are deriving 
proper motions using a subsample
of Dias et al.\ stars and adopting the same astrometry.  A key difference, however, is that 
a number of Dias et al.\ ``member'' stars are excluded by our RV membership criterion
so that, while we typically derive approximately the same bulk motions as Dias et al.\, 
these authors allow many more actual non-members to enter their sample; nevertheless, that Dias et al.\
include more actual non-members seems to have relatively small effects because these authors are typically
averaging over a large number of stars in each cluster, including, apparently, sufficient
numbers of true members to get close to the correct proper motion. 
We show in Table \ref{diascomp} 
the numbers of Dias et al.\ member stars ($P_{Dias} \ge 50$\%) 
that are confirmed to be members (col. 12) and how many we find to be unlikely members (col. 13)
based on the addition of our RV analysis.
On average we find {\it half} of the Dias et al.\ ``member'' 
stars to be non-members when we account for the RVs.
This suggests
that use of proper motion data of the quality of Tycho-2 alone may be insufficient to
determine reliable cluster memberships, though, when averaged over many multiple stars 
and applying the 3$\sigma$ rejection of outlier proper motions adopted by Dias et al.,
these proper motions are useful for deriving the cluster bulk proper motion.  
The Dias et al.\ membership inaccuracies are likely lessened
for closer clusters (e.g., $d < 2$ kpc) which have more bright Tycho-2 stars.
Using the sub-samples from Figure \ref{PMvPMHist},
we see that when both samples have a lot of ``members'' there is convergence to a common proper motion, 
as expected.
We also see that as the sample sizes 
decrease the measured proper motion differences grow. 
It is clear that, at least in our case, 
when we have too few stars 
we may have trouble ``finding'' the true cluster members (e.g., 
as in the examples of NGC 1513 and NGC 7654).
However as both our and the Dias et al.\ also studies drop to a few stars per clusters, 
it is difficult to determine which study is correct.  
We argue that given our more restrictive 3D membership criteria that ours is superior, 
though further study will be needed to confirm this assertion.  
Thus, while Tycho-2 has the best currently available astrometric data, 
more strict RV discrimination such as we provide can substantially improve the
application of these data for determining cluster motions, given a sufficient 
number of RV members.  


\section{Summary}

We have derived high precision (typically $< 3$ km s$^{-1}$ uncertainties)
radial velocities for 3436 stars in the fields of 71 open clusters within 3 kpc of the Sun.
This represents the largest sample of clusters assembled thus far having 
uniformly determined, high-precision radial velocities.  To extend this uniformity to
the other velocity dimensions, 
our survey has focused primarily on obtaining spectra of stars having measured Tycho-2 proper motions;
however, 
our target list was appended with other stars in the cluster fields to expand the membership
census for each cluster.  We have jointly applied three criteria --- spatial position, 
radial velocity and proper motion (in two dimensions) --- to derive high quality cluster membership 
probabilities for the samples stars.   In at least half of our clusters we have found
at least three stars in the field that are reliable members of the cluster using all of these criteria.

Using these member lists, we have averaged the RVs and the Tycho-2 proper motions
to derive mean space velocities for each cluster.  With few exceptions, our mean cluster
RVs are close to those previously derived for the several dozen clusters that have been surveyed 
by other groups.  A comparison of our mean cluster proper motions with those by 
\citet{dias01,dias02a} --- who also relied on Tycho-2 proper motions --- shows
that both data sets are in general agreement, though our results should be more reliable
given our more stringent assessment of cluster membership (i.e., we add
high quality RVs to the proper motion criteria used by Dias et al.).  We find that typically
a large fraction of the Dias et al.\ stars in each cluster field do not meet our most restrictive, joint 
membership criteria.  In a few cases with discrepant proper motion 
results compared to those derived by Dias et al.\,
 we find that the differences may be
due to a critically small numbers of stars surviving our 3D ``membership'' criteria;
i.e. in some of these cases (namely Collinder 258, Lynga 1, NGC 1513, NGC 6250, and NGC 7654) 
it is likely that our results, based on only
one or two stars, might be wrong due to the improper identification of cluster members.
Nevertheless, our data provide reliable 3-D space motions for 66 open clusters.

In most cluster fields we have explored, our 
membership analysis
provides valuable new benchmarks for improved isochrone fitting of the cluster CMDs, which is
useful for estimating ages, distances, metallicities and/or reddenings to these systems.
The resulting
distances and metallicities will allow a new attempt at measuring the Galactic metallicity gradient
with these clusters.  With improved 
distances and more reliable space velocities, the orbits of the clusters can be derived
under an assumed Galactic potential and solar Galactocentric distance.
Alternatively,
these space velocities can be used as tracers of the local velocity field and be used to
investigate the Galactic rotation curve with a set of objects having velocity independent distances 
and uniformly derived, quality space velocities.
We intend to address these science issues in 
future contributions in this series.

Finally, our census of reliable cluster members provides a primary target list for future efforts to 
explore these open clusters with either high resolution spectroscopy or high precision astrometry,
like that expected from SIM PlanetQuest.

\acknowledgments

We are grateful to W.~Butler Burton for useful conversations and Ricardo Mu{\~n}oz 
for discussions and assistance with the WIYN observations.
We thank the anonymous referee for suggestions that helped the presentation of the paper.
We would also like to thank the National Optical Astronomy Observatories (NOAO)
for granting this Ph.D.~dissertation project long-term observing status. 
We acknowledge travel support for PMF from NOAO.
This project was supported by the SIM PlanetQuest key project {\it Taking 
Measure of the Milky Way} under NASA/JPL contract 1228235.  We also
acknowledge funding from 
NSF grant AST-0307851, a David and Lucile Packard Foundation Fellowship to SRM
during the early stages of this project, and the F.H. Levinson 
Fund of the Peninsula Community Foundation.  Additionally, PMF was supported by 
an NSF Astronomy and Astrophysics Postdoctoral Fellowship under award AST-0602221, 
the NASA Graduate Student Researchers Program, a University of Virginia 
Faculty Senate Dissertation-Year Fellowship,   
and grants from the Virginia Space Grant Consortium.
The Tycho-2 catalog is based on observations of the ESA Hipparcos satellite.
This research has made use 
of the USNOFS Image and Catalogue Archive operated by the United States 
Naval Observatory, Flagstaff Station (http://www.nofs.navy.mil/data/fchpix/).
The results presented in this publication also make use of data from
the Two Micron All Sky Survey (2MASS), which is a joint project of the
University of Massachusetts and the Infrared Processing and Analysis
Center (IPAC), funded by the National Aeronautics and Space Administration
and the National Science Foundation.


\begin{deluxetable}{lccrrrccc}
\tabletypesize{\scriptsize}
\tablewidth{0pt}
\tablecaption{\citet{dias02b} Target Cluster Properties\label{cltargets}}
\tablehead{ Cluster & $\alpha_{2000}$ &
 $\delta_{2000}$  &  \multicolumn{1}{c}{$l$ ($\degr$)}  &   \multicolumn{1}{c}{$b$ ($\degr$)} &
 $d($pc$)$  & log($age$) (yr)  &  Diameter($\arcmin$) & \multicolumn{1}{c}{Run} }
\startdata
NGC 129 	 & 00:30:00  & 	$+$60:13:06  & 	120.2701  & 	 1.4566 	 & 1625  & 	7.886  & 	19 &  5 \\
NGC 381 	 & 01:08:19  & 	$+$61:35:00  & 	294.3672  & 	 0.1870 	 & 1148  & 	8.505  & 	6  &  5 \\
NGC 457 	 & 01:19:35  & 	$+$58:17:12  & 	303.2056  & 	 0.2577 	 & 2429  & 	7.324  & 	20 &  5 \\
NGC 884 	 & 02:22:18  & 	$+$57:08:12  & 	141.2419  &  $-$10.6452 	 & 2345  & 	7.032  & 	18 &  5 \\
NGC 957 	 & 02:33:21  & 	$+$57:33:36  & 	124.6487  & $ -$13.4947 	 & 1815  & 	7.042  & 	10 &  5 \\
NGC 1513 	 & 04:09:57  & 	$+$49:30:54  & 	152.5898  &   $-$1.5743 	 & 1320  & 	8.110  & 	10 &  5 \\
NGC 1528 	 & 04:15:23  & 	$+$51:12:54  & 	152.0568  & 	 0.2577 	 &  776  & 	8.568  & 	16 &  5 \\
NGC 1662         & 04:48:27  &  $+$10:56:12  &  187.6949  &  $-$21.1142		 &  437  &      8.625  &        20  &  4 \\
\enddata
\tablenotetext{a}{Run 1: March 2002 (CTIO), Run 2: March 2003 (CTIO), Run 3: June 2003 (CTIO), Run 4: August 2003 (CTIO), Run 5: September 2003 (WIYN)}
\end{deluxetable}

\begin{deluxetable}{lcccccccccc}
\tabletypesize{\scriptsize}
\tablewidth{0pt}
\tablecaption{RV Standard Observations\label{rvstdindiv}}
\tablehead{       &      &               &         &       &      &  $V_{r}$        &  $\epsilon_{V}$  & Avg.         & $\epsilon_{V,avg}$  &IAU \\
         UT Date  & Star & Spec. Type & Frame \# & Fiber &  TDR &\tiny (km s$^{-1}$) & \tiny(km s$^{-1}$)&\tiny (km s$^{-1}$) &\tiny (km s$^{-1}$)&\tiny (km s$^{-1}$)}                 
\startdata 
\multicolumn{11}{c}{Full table given in journal}\\
\enddata
\tablenotetext{a}{Due to poor S/N this star was not used as a cross-corelation template}
\tablenotetext{a}{Not an IAU RV standard star}
\end{deluxetable}

\begin{deluxetable}{lcccccc}
\tabletypesize{\scriptsize}
\tablewidth{0pt}
\tablecaption{Calibration of RV Errors\label{crvsv}}
\tablehead{           &       &Star    & Degrees of &                &           &\\  
 Run Dates & Run\# &Colors  & Freedom    &  $\chi^2_{50}$ & $\alpha$  & Telescope}
\startdata
2002 Mar 10--13      & 1 & Red    &   24       & 23.337         &  35.19  & CTIO 4-m\\
2003 Mar 16--21      & 2 & Red    &   22       & 21.337         &  27.69  & CTIO 4-m\\
2003 July 19--22       & 3 & Red    &   41       & 40.334         &  50.24  & CTIO 4-m\\
2003 Aug 01--07     & 4 & Red    &   72       & 71.333         &  33.59  & CTIO 4-m\\
2003 Aug 01--07     & 4 & Blue\tablenotemark{a} & \phn9      & \phn8.343      &  84.87 & CTIO 4-m\\
\hline\\[-8pt]
2003 Sep 14--17     & 5 & Red    &    \phn6      & \phn5.3481   &  53.22  & WIYN 3.5-m\\
2003 Sep 14--17     & 5 & Blue   &    \phn6      & \phn5.3481   &  41.14  & WIYN 3.5-m\\[-1ex] 
\enddata
\tablenotetext{a}{These data were used as ``blue'' cross-correlation templates for all CTIO runs.}
\end{deluxetable}

\begin{deluxetable}{cccccc}
\tabletypesize{\scriptsize}
\tablewidth{0pt}
\tablecaption{Comparison of RV Standard Star Values\label{rvstdall}}
\tablehead{ Star &  Spectral Type &  UT date & $V_{r}$ (km s$^{-1}$)
&  $V_{r,IAU}$ (km s$^{-1}$) &  Difference (km s$^{-1}$)}
\startdata
HD\phn126053          & G0V      & 11 Mar 2002 &    $-$20.8  $\pm$ 0.6 &     $-$18.5 $\pm$ 0.4 & $-$2.3 $\pm$ 0.7\\
HD\phn150798          & K2II-III & 11 Mar 2002 &\phn$-$1.8   $\pm$ 0.8 & \phn$-$3.7  $\pm$ 0.2 & $+$1.9 $\pm$ 0.8\\
HD\phn157457          & G8III    & 12 Mar 2002 &    $+$17.7  $\pm$ 0.6 &     $+$17.4 $\pm$ 0.2 & $+$0.3 $\pm$ 0.7\\[0.3ex]\hline\\[-6pt]
HD\phn136202          & F8III-IV & 22 Mar 2003 &    $+$54.9  $\pm$ 0.9 &     $+$53.5 $\pm$ 0.2 & $+$1.4 $\pm$ 0.9\\
HD\phn157457          & G8III    & 19 Mar 2003 &    $+$17.5  $\pm$ 0.5 &     $+$17.4 $\pm$ 0.2 & $+$0.1 $\pm$ 0.5\\
HD\phn168454          & K2.5IIIa & 19 Mar 2003 &    $-$20.7  $\pm$ 0.4 &     $-$20.0 $\pm$ 0.0 & $-$0.7 $\pm$ 0.4\\[0.3ex]\hline\\
\enddata
\tablenotetext{a}{Stars HR 675 and HR 7773 are not IAU standards.  Due to the lack of IAU standards hotter than F type stars, we used stars from \citet{fekel}.}
\end{deluxetable}

\begin{deluxetable}{lcccc}
\tabletypesize{\scriptsize}
\tablewidth{0pt}
\tablecaption{Comparison of Measured Red vs. Blue RVs \label{fitRB}}
\tablehead{ Run  &  $a_0$  &  $a_1$  & RMS & Blue Correction \\
                 &         &         &     & (km s$^{-1}$)  }
\startdata
March 2002            &  $-$0.47 & 0.97 & 3.06  & \phn\phs0.0 \\
March 2003            &  $-$1.26 & 0.99 & 3.74  &  \phn$-$1.3\\
August 2003           & $-$10.09 & 1.00 & 4.43  &  $-$10.0\\
September 2003        &  $-$6.05 & 0.98 & 4.63  &  \phn$-$6.0\\[-1ex]
\enddata
\end{deluxetable}
\renewcommand{\arraystretch}{1.0}

\begin{deluxetable}{ l c c c c c c r c r c c} \tabletypesize{\small}
\tablewidth{0pt}
\tabletypesize{\tiny}
\tablecaption{NGC 2682 Comparison to Previous RV Results \label{clM67comp}}
\tablehead{
& & & & &  
 $\alpha_{2000}$ &
 $\delta_{2000}$ &
 \multicolumn{1}{c}{$V_{r}$} & 
 $\epsilon_{V_{r}}$ &  \multicolumn{1}{c}{$V_{r, Mathieu}$} & 
 $\epsilon_{V_{r}}$ & $\Delta V_{r}$\\
 Star (Tycho-2)& SAND & FBC & MMJ & ES &{\scriptsize (hr)} &{\scriptsize (\degr)} &
 {\tiny (km s$^{-1}$)} &
 {\tiny (km s$^{-1}$)} &
 {\tiny (km s$^{-1}$)} &
 {\tiny (km s$^{-1}$)} &
 {\tiny (km s$^{-1}$)} } 
\startdata
     0813-01125-1 &   0721 &   2554  &  6475 & \nodata &  8:50:49.65 &  11:35:08.9 &     $+$33.01 &  0.79 & $+$34.2 & 0.9   &  $-$1.2 \\
     0813-02302-1 &   0727 &   2425  &  5158 & \nodata &  8:50:44.98 &  11:37:30.4 &     $+$21.44 &  1.82 & $+$16.3 &\nodata&  \phs5.1 \\
     0814-02331-1 &   0978 &   3364  &  6482 & \nodata &  8:51:17.48 &  11:45:22.7 &     $+$35.17 &  0.59 & $+$34.7 & 0.6   &  \phs0.5 \\
     0814-01531-1 &   1010 &   3536  &  6485 &   1008  &  8:51:22.81 &  11:48:01.8 &     $+$33.00 &  0.69 & $+$33.6 & 0.4   & $-$0.6 \\
     0814-01493-1 &   1016 &   3349  &  6484 & \nodata &  8:51:17.10 &  11:48:16.2 &     $+$35.66 &  0.59 & $+$34.3 & 0.7   &  \phs1.4 \\
     0814-01631-1 &   1023 &   3668  &  6487 &   1001  &  8:51:26.84 &  11:48:40.5 &      $+$3.54 &  1.01 &  $+$3.6 & 1.0   & $-$0.1 \\
     0814-01119-1 &   1040 &   3566  &  6488 & \nodata &  8:51:23.78 &  11:49:49.4 &     $+$34.60 &  0.82 & $+$32.0 & 6.3   &  \phs2.6 \tablenotemark{a}\\
     0814-01763-1 &   1054 &   3347  &  6489 &   4020  &  8:51:17.03 &  11:50:46.4 &     $+$32.63 &  0.72 & $+$33.5 & 0.4   & $-$0.9 \\
\enddata
\tablerefs{SAND: \citet{sand77}, FBC: \citet{fbc}, MMJ: \citet{mmj93},
ES: \citet{es64}, Mathieu: \citet{mathieu86}.}
\tablenotetext{a}{Star listed at spectroscopic binary in \citet{mathieu86}.}
\end{deluxetable}

\begin{deluxetable}{ l r r r r r r c} \tabletypesize{\small}
\tablewidth{0pt}
\tabletypesize{\tiny}
\tablecaption{Star-by-Star Comparison to Previous Results \label{clcomp}}
\tablehead{  & \multicolumn{1}{c}{$\alpha_{2000}$}& \multicolumn{1}{c}{$\delta_{2000}$}&\multicolumn{1}{c}{$V_{r}$} & \multicolumn{1}{c}{$V_{r}$ (Other)} &  \multicolumn{1}{c}{$\Delta V_r$}   & Other &  Other \\
        Star & \multicolumn{1}{c}{(hr)}           & \multicolumn{1}{c}{(\degr)}        & \multicolumn{1}{c}{(km s$^{-1}$)}&\multicolumn{1}{c}{(km s$^{-1}$)}& \multicolumn{1}{c}{(km s$^{-1}$)} &   ID  &  Ref.  }
\startdata
\multicolumn{8}{c}{Full table given in journal}\\
\enddata
\tablerefs{ 1: \citet{mermio95}, 2: \citet{meibom02}, 3: \citet{mm90}, 4: \citet{mermio96}, 5: \citet{mm89}, 6: \citet{mermio92} }
\tablenotetext{a}{Star listed at spectroscopic binary.}
\end{deluxetable}

\clearpage
\begin{deluxetable}{lcccrcc}
\tabletypesize{\scriptsize}
\tablewidth{0pt}
\tablecaption{Statistics of Observed Cluster Stars \label{fullsample}}
\tablehead{ Cluster  & UT-Date & Exposures  & $N_{Dias}$ &\# Observed\tablenotemark{a} & \# $V_r$ Stars \\
                     &         &            &            &(total/TYC)                  & (total/TYC)      }
\startdata
  NGC 129        & 2003 Sep 15   & $3 \times 600s$ &  29  &  67/\phn29  &\phn 29/29 \\
  NGC 381        & 2003 Sep 15   & $3 \times 600s$ &  20  &  65/\phn20  &\phn 20/20 \\ 
  NGC 457        & 2003 Sep 17   & $3 \times 600s$ &  17  &  61/\phn17  &\phn 17/17 \\
  NGC 884        & 2003 Sep 18   & $3 \times 600s$ &  23  &  45/\phn23  &\phn 23/23 \\
  NGC 957        & 2003 Sep 16   & $3 \times 600s$ &  19  &  63/\phn19  &\phn 19/19 \\ 
  NGC 1513       & 2003 Sep 15   & $3 \times 600s$ &  19  &  48/\phn19  &\phn 19/19 \\
  NGC 1528       & 2003 Sep 17   & $3 \times 600s$ &  43  &  43/\phn43  &\phn 43/43 \\
  NGC 1662       & 2003 Aug 07   & $3 \times 600s$ &  13  & 108/\phn13  &\phn 70/13 \\
\enddata
\tablenotetext{a}{$N_{Dias}$ is the number of Tycho-2 stars observed, which are taken from the study of \citet{dias01,dias02a}.}
\end{deluxetable}

\begin{deluxetable}{ l c c r r r c r r r r}
\tabletypesize{\tiny}
\tablewidth{0pt}
\tablecaption{ Cluster Analysis -- NGC 2682 Example\tablenotemark{a} \label{table:NGC_2682}}
\tablehead{ &  $\alpha_{2000}$ &
 $\delta_{2000}$ &
 \multicolumn{1}{c}{$\mu_{\alpha} \cos \delta$} & 
 \multicolumn{1}{c}{$\mu_{\delta}$}  &
 \multicolumn{1}{c}{$V_{r}$} & &
 \multicolumn{4}{c}{Probabilities} \\\cline{8-11}
 Star &{\tiny (hr)} &{\tiny (\degr)} &
 \multicolumn{1}{c}{\scriptsize (mas yr$^{-1}$)} &
 \multicolumn{1}{c}{\scriptsize (mas yr$^{-1}$)} &
 \multicolumn{1}{c}{\scriptsize (km s$^{-1}$)} &
Type\tablenotemark{b}&\multicolumn{1}{c}{Dias} & $P_c^K$&$P_c^V$& $P_c^{tot}$ }
\startdata
 f9754-MMJ5132  &  8:50:42.52 &      11:39:49.7 &                      \nodata &                    \nodata       &        33.53 $\pm$       3.24 &     Red &        c &      0.0 &      100.0 &      0.0 \\ 
 0813-01521-1   &  8:50:43.57 &      11:35:48.8 &       25.9 $\pm$        1.7 &    $-$32.7 $\pm$        1.9   &      $-$3.84 $\pm$       1.07 &     Red &       0 &      0.0 &      0.0 &      0.0 \\ 
 0813-02302-1   &  8:50:44.98 &      11:37:30.4 &     $-$1.5 $\pm$        2.2 &     $-$2.2 $\pm$        2.5   &        21.44 $\pm$       1.82 &     Red &       0 &        1.3 &        4.5 &      0.0 \\ 
 H800-FBC2476   &  8:50:46.82 &      11:52:57.4 &                      \nodata &                    \nodata       &     $-$76.71 $\pm$       5.47 &     Red &        c &      0.0 &      0.0 &      0.0 \\ 
\bf 0813-01125-1 & \bf 8:50:49.65 & \bf  11:35:08.9 & \bf $-$6.5 $\pm$ \bf    1.6 & \bf $-$5.4 $\pm$ \bf  1.6 & \bf    33.01 $\pm$ \bf   0.79 & \bf Red & \bf  98 & \bf   87.9 & \bf   99.3 & \bf   87.3 \\ 
 H822-FBC2800   &  8:50:58.41 &      11:58:14.4 &                      \nodata &                    \nodata       &        31.22 $\pm$       4.30 &     Red &        c &      0.0 &       88.7 &      0.0 \\ 
 0813-02294-1   &  8:51:03.51 &      11:45:02.7 &     $-$8.8 $\pm$        1.9 &     $-$7.2 $\pm$        1.8   &        38.36 $\pm$       2.04 &     Red &      99 &       72.1 &       63.1 &       45.5 \\ 
 H801-FBC3002   &  8:51:05.82 &      11:53:10.7 &                      \nodata &                    \nodata       &        39.19 $\pm$       4.44 &     Red &        c &      0.0 &       53.3 &      0.0 \\
\enddata
\tablenotetext{a}{Boldface entries denote star selected as cluster members using the 3D criteria.}
\tablenotetext{b}{Type of spectral template used in the cross-correlation for RV determination, as described in \S \ref{stddescrip}.}
\tablenotetext{c}{Star not used in \citet{dias01} membership analysis.}
\end{deluxetable}

\begin{deluxetable}{ l c c r r r c r r r r}
\tabletypesize{\tiny}
\tablewidth{0pt}
\tablecaption{ Cluster Analysis -- All data\tablenotemark{a}\label{alldata}}
\tablehead{ &  $\alpha_{2000}$ &
 $\delta_{2000}$ &
 \multicolumn{1}{c}{$\mu_{\alpha} \cos \delta$} & 
 \multicolumn{1}{c}{$\mu_{\delta}$}  &
 \multicolumn{1}{c}{$V_{r}$} & &
 \multicolumn{4}{c}{Probabilities} \\\cline{8-11}
 Star &{\tiny (hr)} &{\tiny (\degr)} &
 \multicolumn{1}{c}{\scriptsize (mas yr$^{-1}$)} &
 \multicolumn{1}{c}{\scriptsize (mas yr$^{-1}$)} &
 \multicolumn{1}{c}{\scriptsize (km s$^{-1}$)} &
Type\tablenotemark{b} &\multicolumn{1}{c}{Dias} & $P_c^K$&$P_c^V$& $P_c^{tot}$ }
\startdata
\multicolumn{11}{c}{Full table given in journal}\\
\enddata
\tablenotetext{a}{Boldface entries denote star selected as cluster members using the 3D criteria}
\tablenotetext{b}{Type of spectral template used in the cross-correlation for RV determination, as described in \S \ref{stddescrip}}
\tablenotetext{c}{Star not used in \citet{dias01} membership analysis.}
\tablenotetext{d}{Star not analyzed due to large errors (see \S \ref{datacuts}).}
\end{deluxetable}

\begin{deluxetable}{ l l l r r r r r r c}
\tabletypesize{\tiny}
\tablewidth{0pt}
\tablecaption{2MASS and Tycho-2 Photometry for Targeted Cluster Stars\label{ALLPHOT}}
\tablehead{Star & \multicolumn{1}{c}{2MASS ID} &\multicolumn{1}{c}{$\alpha_{2000}$} & \multicolumn{1}{c}{$\delta_{2000}$} & \multicolumn{1}{c}{$B_{Tycho}$} & \multicolumn{1}{c}{$V_{Tycho}$} & \multicolumn{1}{c}{2MASS $J$} & \multicolumn{1}{c}{2MASS $H$} & \multicolumn{1}{c}{2MASS $K_s$} & \multicolumn{1}{c}{3D Memb?} }
\startdata
\multicolumn{9}{c}{Full table given in journal}\\
\enddata
\tablenotetext{a}{Star not matched to any 2MASS point source.}
\end{deluxetable}


\begin{deluxetable}{lrrrrrrrr}
\tabletypesize{\tiny}
\tablewidth{0pt}
\tablecaption{Derived Cluster Bulk Kinematics\label{bulkALL}}
\tablehead{         &\multicolumn{2}{c}{Members} & \multicolumn{1}{c}{Bulk $V_{r,3D}$}    & \multicolumn{1}{c}{Bulk $V_{r,3D+RV}$}    & \multicolumn{1}{c}{$\mu_{\alpha \cos \delta}$}  & \multicolumn{1}{c}{$\mu_{\delta}$}  & \multicolumn{1}{c}{$\mu_{l \cos b}$}  & \multicolumn{1}{c}{$\mu_{b}$}   \\
Cluster  &  3D  & 3D+RV                  & \multicolumn{1}{c}{\tiny (km s$^{-1}$)} & \multicolumn{1}{c}{\tiny (km s$^{-1}$)} & \multicolumn{1}{c}{\tiny (mas yr$^{-1}$)} & \multicolumn{1}{c}{\tiny (mas yr$^{-1}$)} & \multicolumn{1}{c}{\tiny (mas yr$^{-1}$)} & \multicolumn{1}{c}{\tiny (mas yr$^{-1}$)} }
\startdata
     Berkeley 86 &  2 &  - & $ -25.54 \pm    2.60$  &    \nodata             & $  -3.80 \pm    1.27$  & $  -4.58 \pm    1.24$  & $  -5.88 \pm 1.25$ & $   0.90 \pm 1.26$  \\
   Collinder 205 &  1 &  6 & $  27.40 \pm    4.07$  & $  28.18 \pm    1.52$  & $  -4.00 \pm    2.10$  & $   8.00 \pm    1.90$  & $  -8.08 \pm 1.90$ & $  -3.84 \pm 2.10$  \\
   Collinder 258 &  2 &  6 & $  16.04 \pm    0.60$  & $  14.90 \pm    0.93$  & $  -0.90 \pm    2.62$  & $  -5.00 \pm    2.40$  & $   0.35 \pm 2.61$ & $  -5.07 \pm 2.41$  \\
      Harvard 10 &  3 &  - & $ -18.17 \pm    1.94$  &    \nodata             & $  -3.88 \pm    1.07$  & $ -11.64 \pm    1.11$  & $ -11.99 \pm 1.11$ & $   2.58 \pm 1.07$  \\
         IC 2488 &  3 &  - & $  -1.84 \pm    1.52$  &    \nodata             & $  -7.73 \pm    1.56$  & $   8.93 \pm    1.50$  & $  -9.84 \pm 1.50$ & $  -6.55 \pm 1.56$  \\
         IC 4651 &  9 & 10 & $ -33.34 \pm    0.67$  & $ -33.30 \pm    0.67$  & $  -1.72 \pm    0.70$  & $  -2.76 \pm    0.69$  & $  -2.54 \pm 0.69$ & $   2.04 \pm 0.70$  \\
         IC 4756 &  7 & 13 & $ -24.44 \pm    0.75$  & $ -25.08 \pm    0.64$  & $   0.68 \pm    0.74$  & $  -1.99 \pm    0.72$  & $  -1.95 \pm 0.72$ & $  -0.76 \pm 0.74$  \\
    Kharchenko 1 &  3 &  - & $  12.74 \pm    1.31$  &    \nodata             & $   6.10 \pm    1.23$  & $  -1.91 \pm    1.29$  & $   3.54 \pm 1.29$ & $   5.32 \pm 1.23$  \\
\enddata
\tablenotetext{a}{Bulk cluster parameters unreliable due to mmbership uncertainty.}
\end{deluxetable}

\begin{deluxetable}{l c c c c c c c c c} \tabletypesize{\small}
\tablewidth{0pt}
\tabletypesize{\small}
\tablecaption{Derived NGC 2682 Motions Compared to Previous Results\label{clcomp2}}
\tablehead{ 
 Parameter & Units & \multicolumn{1}{c}{$N_{members}$} & \multicolumn{1}{c}{Value} & Ref}
\startdata
$\mu_{\alpha} \cos \delta$  &(mas yr$^{-1}$) &     10             & $-7.87 \pm 0.61$  &    This Work\\  
                            &(mas yr$^{-1}$) &     30             & $-8.62 \pm 0.28$  &  \citet{dias01}       \\
                            &(mas yr$^{-1}$) &     27             & $-8.31 \pm 0.26$  &    \citet{khar05}     \\\hline
$\mu_{\delta}$              &(mas yr$^{-1}$) &     10             & $-5.60 \pm 0.59$  &  This Work  \\  
                            &(mas yr$^{-1}$) &     30             & $-6.00 \pm 0.28$  &  \citet{dias01}       \\
                            &(mas yr$^{-1}$) &     27             & $-4.81 \pm 0.22$  &   \citet{khar05}      \\\hline
$V_{r}$                     &(km s$^{-1}$)   &     10             &  $33.67 \pm  0.42$ &   This Work \\ 
                            &(km s$^{-1}$)   &     13             &  $ 32   \pm 9$     &  \citet{sfj95}       \\
                            &(km s$^{-1}$)   &    104             &  $33.5 \pm  0.5$ &    \citet{mathieu86}   \\[-1ex]   
\enddata
\end{deluxetable}

\begin{deluxetable}{lccrrrcrrl}
\tabletypesize{\tiny}
\tablewidth{0pt}
\tablecaption{Comparison of Derived Cluster RVs \label{bulkCompare}}
\tablehead{         &\multicolumn{2}{c}{Members} & \multicolumn{1}{c}{Bulk $V_{r,3D}$}   & \multicolumn{1}{c}{Bulk $V_{r,3D+RV}$}  & \multicolumn{1}{c}{Other $V_r$} & \# & $\Delta V_{r,3D}$ & $\Delta V_{r,3D+RV}$ & \\
Cluster  &  3D  & 3D+RV                  & \multicolumn{1}{c}{\tiny (km s$^{-1}$)} & \multicolumn{1}{c}{\tiny (km s$^{-1}$)} & \multicolumn{1}{c}{\tiny (km s$^{-1}$)} & stars & {\tiny (km s$^{-1}$)}& {\tiny (km s$^{-1}$)}& Reference }
\startdata
         Berkeley 86 &  2 &  - & $-$25.54 $\pm$  2.64 &     \nodata          &  $-$22.0  $\pm$  7.0   &   6    &  $+$3.5 & \nodata &   \citet{forbes92}\\[1.5ex]
             IC 2488 &  3 &  - &  $-$1.84 $\pm$  1.52 &     \nodata          &   $-$2.6  $\pm$  0.1   &\nodata &  $-$0.8 & \nodata &   \citet{cmpm03} \\[1.5ex]
             IC 4651 &  9 & 10 & $-$33.34 $\pm$  0.67 & $-$33.30 $\pm$ 0.67  &  $-$31.0  $\pm$  0.2   &   14   &  $+$2.3 &  $+$2.3 &   \citet{mermio95}\\
                     &    &    &                      &                      &  $-$30.8  $\pm$  0.3   &   44   &  $+$2.5 &  $+$2.5 &   \citet{meibom02}\\[1.5ex]
             IC 4756 &  7 & 13 & $-$24.44 $\pm$  0.75 & $-$25.08 $\pm$  0.64 &  $-$25.8  $\pm$  0.6   &   13   &  $-$1.4 &  $-$0.7 &   \citet{mm90} \\
                     &    &    &                      &                      &  $-$25.0  $\pm$  0.2   &   15   &  $-$0.6 &  $+$0.1 &   \citet{val90}\\[1.5ex]
             NGC 129 &  3 &  - & $-$39.41 $\pm$  0.54 &     \nodata          &  $-$38.5  $\pm$  0.2   &    2   &  $+$0.9 & \nodata &  \citet{mermio87} \\[1.5ex]
             NGC 457 &  2 &  - & $-$32.99 $\pm$  0.60 &     \nodata          &  $-$25.1  $\pm$  3.0   &    4   &  $+$7.9 & \nodata &   \citet{li89}\\
\enddata
\end{deluxetable}

\begin{deluxetable}{lrrrrrrrrrrrrr} 
\tabletypesize{\tiny}
\tablewidth{0pt}
\tablecaption{Comparison of Derived Mean Proper Motions to those of \citet{dias01,dias02a}\label{diascomp}}
\tablehead{          & \multicolumn{4}{c}{\citet{dias01,dias02a}}                        && \multicolumn{3}{c}{This Study}        & & \multicolumn{4}{c}{Comparison}     \\\cline{2-5}\cline{7-9}\cline{11-14}
                     &            &       & \multicolumn{1}{c}{$\mu_{\alpha \cos \delta}$} & \multicolumn{1}{c}{$\mu_{\delta}$}  &&  \multicolumn{1}{c}{}               & \multicolumn{1}{c}{$\mu_{\alpha \cos \delta}$} & \multicolumn{1}{c}{$\mu_{\delta}$}  & &   $\Delta \mu_{\alpha \cos \delta}$  &   \multicolumn{1}{c}{$\Delta \mu_{\delta}$} & \multicolumn{2}{c}{Dias Memb.} \\\cline{13-14}
            Cluster  &    Stars   & Memb. & {\tiny (mas yr$^{-1}$)} & {\tiny (mas yr$^{-1}$)} & & Memb   & {\tiny (mas yr$^{-1}$)} & {\tiny (mas yr$^{-1}$)} &&{\tiny (mas yr$^{-1}$)} & {\tiny (mas yr$^{-1}$)}   & Conf. & Rej. }
\startdata
Berkeley 86               &   99 &   50 &$  -4.1 \pm 2.2 $&$  -4.5 \pm 2.2 $&&   2 &$  -3.8 \pm 1.3 $&$  -4.6 \pm 1.2 $&&$  -0.3 $&$   0.1 $&   2 &  22   \\
Collinder 205             &   19 &   12 &$  -3.9 \pm 1.9 $&$   6.5 \pm 1.9 $&&   1 &$  -4.0 \pm 2.1 $&$   8.0 \pm 1.9 $&&$   0.1 $&$  -1.5 $&   1 &   4   \\ 
Collinder 258\tablenotemark{a}& 25 & 13 &$  -8.1 \pm 2.2 $&$  -0.8 \pm 2.2 $&&   2 &$  -0.9 \pm 2.6 $&$  -5.0 \pm 2.4 $&&$  -7.2 $&$   4.2 $&   0 &   8   \\
Harvard 10                &   92 &   34 &$  -2.8 \pm 1.6 $&$ -11.2 \pm 1.6 $&&   3 &$  -3.9 \pm 1.1 $&$ -11.6 \pm 1.1 $&&$   1.1 $&$   0.4 $&   3 &  11   \\
IC 2488                   &   63 &   40 &$  -5.6 \pm 3.0 $&$   8.0 \pm 3.0 $&&   3 &$  -7.7 \pm 1.6 $&$   8.9 \pm 1.5 $&&$   2.1 $&$  -0.9 $&   3 &  22   \\
IC 4651                   &   43 &   19 &$  -1.1 \pm 2.1 $&$  -2.2 \pm 2.1 $&&   9 &$  -1.7 \pm 0.7 $&$  -2.8 \pm 0.7 $&&$   0.6 $&$   0.6 $&   7 &  16   \\
IC 4756                   &  181 &   30 &$  -0.1 \pm 1.3 $&$  -3.4 \pm 1.3 $&&   7 &$   0.7 \pm 0.7 $&$  -2.0 \pm 0.7 $&&$  -0.8 $&$  -1.4 $&   4 &   8   \\
Kharchenko 1              &   86 &   40 &$   2.1 \pm 3.6 $&$  -3.5 \pm 3.6 $&&   3 &$   6.1 \pm 1.2 $&$  -1.9 \pm 1.3 $&&$  -4.0 $&$  -1.6 $&   3 &  14   \\
\enddata
\tablenotetext{a}{Cluster excluded from further analysis, due to  $\Delta \mu_{\alpha \cos \delta}$ or  $\Delta \mu_{\delta}  > 5.0$ mas yr$^{-1}$.}
\end{deluxetable}

\clearpage


\clearpage

\end{document}